\documentclass[a4paper,11pt]{article}
\pdfoutput=1 

\usepackage{jheppub} 

\usepackage[T1]{fontenc} 
\usepackage[pdftex]{pict2e}
\usepackage{tikz}

\title{\boldmath Holographic collisions in large $D$ effective theory}

\author[1]{Raimon~Luna,}
\author[2]{Mikel~Sanchez-Garitaonandia}

\affiliation[1]{Departamento de Astronom\'{i}a y Astrof\'{i}sica, Universitat de Val\`{e}ncia, Dr. Moliner 50, 46100, Burjassot (Val\`{e}ncia), Spain}
\affiliation[2]{CPHT, CNRS, \'Ecole polytechnique, Institut Polytechnique de Paris, 91120 Palaiseau, France}

\emailAdd{raimon.luna-perello@uv.es}
\emailAdd{mikel.sanchez@polytechnique.edu}

\abstract{We study collisions of Gaussian mass-density blobs in a holographic plasma, using a large $D$ effective theory, as a model for holographic shockwave collisions. The simplicity of the effective theory allows us to perform the first 4+1 collisions in Einstein-Maxwell theory, which are dual to collisions of matter with non-zero baryonic number. We explore several collision scenarios with different blob shapes, impact parameters and charge values and find that collisions with impact parameter below the transverse width of the blobs are equivalent under rescaling. We also observe that charge weakly affects the rest of quantities. Finally, we study the entropy generated during collisions, both by charge diffusion and viscous dissipation. Multiple stages of linear entropy growth are identified, whose rates are not independent of the initial conditions.}

\begin{document} 

\begin{flushright}
CPHT-RR088.122022
\end{flushright}
\maketitle
\flushbottom

\section{Introduction}
\label{sec:Introduction}
The AdS/CFT duality has shed light on some important, and hardly accessible otherwise, features of strongly coupled, out-of-equilibrium quantum systems. Particularly, a very relevant field of application has been the Heavy Ion Collision program, where the collision of high energy shockwaves in AdS are used as a proxy for colliding ions in a particle accelerator and collider. Several studies have been carried out in this direction, from the study of particular aspects of the collision dynamics \cite{Nastase:2005rp, Janik:2005zt, Janik:2006gp, Kovchegov:2007pq, Grumiller:2008va, Lin:2009pn, Beuf:2009cx, Kovchegov:2009du, Gubser:2009sx, Romatschke:2013re, Bantilan:2018vjv} to simplified modelling of the full collision \cite{Kajantie:2008rx, Albacete:2009ji}. Moreover, fully numerical simulations of shockwave collisions were done in pure AdS$_5$ \cite{Chesler:2010bi, Casalderrey-Solana:2013aba, Chesler:2015wra, vanderSchee:2015rta, Grozdanov:2016zjj, Waeber:2019nqd, Muller:2020ziz, Waeber:2022tts, Waeber:2022vgf}, and in non-conformal quantum field theories \cite{Attems:2016tby, Attems:2017zam} including those with thermal phase transitions \cite{Attems:2018gou}. Additionally, a non-vanishing baryonic density was included in \cite{Casalderrey-Solana:2016xfq}.

While the numerical simulation of these gravitational phenomena is very much possible, general shockwave collisions are computationally expensive as they imply numerically solving Einstein's equations, ideally with a 4+1 dependence, with no symmetry assumptions. A first simplification to be done is to reduce the number of dynamical dimensions, such as the collision of planar shocks \cite{Chesler:2010bi}, which is justified by the high Lorentz contraction of rapidly moving ions. This reduces the problem to 2+1 dimensions and has proved to be useful to gain insight into the collision dynamics. Going a bit beyond, \cite{Waeber:2022tts} has recently proposed to perform an expansion in gradients transverse to the collision direction. This has revealed that even the first nontrivial order is able to capture a surprisingly high amount of physics involved in a full AdS$_5$ shockwave collision \cite{Chesler:2015wra}. By means of this approach, a collision with a more realistic model for the energy distribution inside the nuclei was first performed in \cite{Waeber:2022vgf}.

Here we propose to tackle the problem with a different kind of approximation by taking the limit of a very large number of spacetime dimensions, which is known as the large $D$ limit of General Relativity \cite{Emparan:2013moa} (see \cite{Emparan:2020inr} for a review). In this limit, the horizon dynamics gets decoupled from the region far from the horizon and, as a result, gravitational waves are decoupled from the dynamics of the horizon \cite{Emparan:2014cia, Emparan:2014aba}. The resulting effective description represents a major simplification \cite{Emparan:2015hwa} with respect to Einstein's equations at finite dimension $D$. Such a limit has been already useful in studying many properties of black holes \cite{Emparan:2013xia, Emparan:2013oza, Emparan:2015rva, Emparan:2019obu} and the effective theory has been widely used to understand the classical dynamics of horizons: instabilities, turbulent behavior and even violation of the Weak Cosmic Censorship conjecture in both asymptotically flat and AdS spacetimes \cite{Emparan:2014jca, Emparan:2015gva, Emparan:2016sjk, Rozali:2017bll, Andrade:2018nsz, Andrade:2018rcx, Andrade:2018yqu, Andrade:2018zeb, Andrade:2019edf, Andrade:2019rpn, Andrade:2020ilm, Emparan:2021ewh}.

In the present work will we will focus on 4+1-dimensional holographic collisions of blobs in Einstein-Maxwell theory, which adds a non-vanishing baryonic number density on the boundary theory \cite{Emparan:2016sjk}. Recall that 4+1 holographic collisions are dual to 3+1-dimensional collisions in the boundary conformal quantum field theory. This means that we will take only $4$ spatial dimensions out of the infinite number of them to be non passive. The effect of the spectator dimensions is to dilute the gravitational field, strongly focusing it near the horizon. The resulting effective description is non-relativistic, it has different transport coefficients, and the background horizon temperature only differs from that of the blobs by $1/D$ corrections. This last point, which is possibly the most relevant, implies that dissipation of the initial blobs cannot be arbitrarily suppressed. 

However, the simplification of the description means that the computational cost of evolving its equations of motion is small. This fact allows us to scan over parameters in order to obtain a qualitative picture of the possible differences that arise during the collisions. Questions about the importance of the baryonic density, and the dependency of the results on the impact parameter can be therefore addressed. Additionally, we are able of reaching further in time during collisions than in the past \cite{Chesler:2015wra}.

In particular, we are interested in studying the production of entropy during the collisions. In the past, linear growths in time of the total entropy were observed in the context of AdS$_5$ shock collisions, see e.g. \cite{Grozdanov:2016zjj,Muller:2020ziz}. We wonder if such behavior will be captured by the large $D$ effective theory and if, as it was claimed in \cite{Muller:2020ziz}, we can link the growth rate to Lyapunov exponents. We will see that scanning over different initial data setups we can stablish the possible sensitivity of the growth rate to such details in order to determine whether there is a connection with chaotic behavior.

The paper is organized as follows: In Section \ref{sec:Large-D} we introduce the large $D$ effective equations of motion and the setup. We then give an overview of the collisions in Section \ref{sec:Collisions}, including how the evolution changes with the impact parameter, and to what extent the baryonic charge plays a role. We then focus on the entropy growth in Section \ref{sec:Entropy}, for both vanishing and non-vanishing charge density. We finally conclude in Section \ref{sec:Discussion}.

\section{The large $D$ effective equations}
\label{sec:Large-D}

Let us consider a gravitational theory dual to a conformal theory with the addition of a $U(1)$ global symmetry playing the role of a baryon number. The action of such theory reads
\begin{equation}
    I = \int d^{D}x \sqrt{-g}\left(R-\frac{1}{4}F^2-2\Lambda\right),
\end{equation}
with $\Lambda=-n(n-1)/2$ the cosmological constant, $n=D-1$ and $F_{\mu\nu}$ the Maxwell field strength tensor. Throughout the paper we will be working in units in which $16\pi G =\Omega^{n+1}$, the area of the $n+1$-dimensional unit sphere.

We now follow the same steps from \cite{Emparan:2016sjk} to obtain the large $D$ effective equations. We start by observing that, as $n$ increases, the speed of sound scales as $c_s=1/\sqrt{n-1}$, and so the theory will become non-relativistic. In order to capture the arising physics we need to focus on velocities and distances of $\mathcal{O}(1/\sqrt{n})$, so we rescale our spatial coordinates and $g_{it}$ by $1/\sqrt{n}$ in order to work with order one quantities\footnote{We will write the physical quantities with boldface while the rescaled ones in regular typography, e.g. $\boldsymbol{v^i} = v^i/\sqrt{n}$ and $\boldsymbol{x}^i=x^i/\sqrt{n}$}. A general AdS black brane geometry in ingoing Eddington-Finkelstein coordinates reads, under these rescalings,
\begin{equation}
    ds^2 = r^2\left(-Adt^2-\frac{2}{n}C_idtdx^i+\frac{1}{n}G_{ij}dx^idx^j\right)-2dtdr,
\end{equation}
where $r$ is the holographic coordinate and $x_i$ are the rescaled, order one coordinates along the horizon with $i=1,2..., n-1$. The factors of $1/n$ result from the rescalings. Furthermore, if we want the gauge field to backreact on the metric at leading order in $1/n$ we need to take $A_t = \mathcal{O}(1)$ and $A_i = \mathcal{O}(1/n)$. Upon substitution in Einstein's equations, one can solve them as a series expansion in $1/n$. To leading order, the result is,
\begin{equation}
\begin{aligned}
    A & = 1- \rho(t,\vec x)\left(\frac{r_0}{r}\right)^n+q(t,\vec x)^2\left(\frac{r_0}{r}\right)^{2n},\\
    C_i & = p_i(t,\vec x)\left(\frac{r_0}{r}\right)^n\left(1-q(t,\vec x)^2\left(\frac{r_0}{r}\right)^{2n}\right),\\
    G_{ij} & = \delta_{ij}+\frac{1}{n}\left[\frac{C_ip_j(t,\vec x)}{\rho(t,\vec x)}-\log\left(1-\rho_-(t,\vec x)\left(\frac{r_0}{r}\right)^n\right)\partial_{(i}\left(\frac{p_{j)}(t,\vec x)}{\rho(t,\vec x)}\right)\right],\\
\end{aligned}
\end{equation}
where $r_0$ indicates the position of the unperturbed, neutral horizon (corresponding to $\rho=1$ and $q=p_i=0$). The remaining variables, $\rho=\boldsymbol{\rho}/n$, $q=\boldsymbol{q}/n$ and $p^i=\boldsymbol{p}^i/n$ are the mass, charge and momentum densities respectively. From now on, we will set $r_0^n=1$ which fixes the units in which we will work. The quantities $\rho_{\pm}$ are defined as,
\begin{equation}
    \rho_{\pm} = \frac{1}{2}\left(\rho\pm\sqrt{\rho^2-2q^2}\right),
\end{equation}
and the horizon is located at $r^n = \rho_+$. To next order in the $1/n$ expansion, one obtains the equations of motion for the new variables,
\begin{equation}
\begin{aligned}
    \partial_t\rho-\partial_i\partial^i\rho+\partial_ip^i & = 0,\\
    \partial_tq-\partial_i\partial^iq+\partial_i\left(\frac{p^iq}{\rho}\right) & = 0,\\
    \partial_tp_i-\partial_j\partial^jp_i+\partial_i\rho+\partial^j\left[\frac{p_ip_j}{\rho}+\rho_-\left(\partial_i\frac{p_j}{\rho}+\partial_j\frac{p_i}{\rho}\right)\right] & = 0.
\end{aligned}
\label{eq:Large-D-effecttive-equations}
\end{equation}
Notice that, in order for $\rho_\pm$ to be real, we need to have $\rho \leq \sqrt{2}q$, where the extremal limit saturates the inequality. The extremal solution should get rid of the high normal gradients ($T=0$), and so we expect the effective theory to break down and the equations to cease being valid. The ratio $\sqrt{2}q/\rho$, ranging from $0$ to $1$, will be later used as a measure of extremality. We can obtain the thermodynamic quantities in the usual manner, as
\begin{equation}
    s = \boldsymbol{s} = 4\pi\rho_+, \quad T = \frac{\boldsymbol{T}}{n} = \frac{\rho_+-\rho_-}{4\pi\rho_+}, \quad \mu = \boldsymbol{\mu}= \frac{q}{\rho_+}. 
    \label{eq:Thermo}
\end{equation}
For illustration purposes, a more familiar set of equations can be obtained by performing the change of variables $p_i = \partial_i\rho + \rho v_i$, by which the equations of motion take the form
\begin{equation}
\begin{aligned}
        \partial_t\rho+\partial_i\left(\rho v^i\right) &= 0,\\
        \partial_tq+\partial_i j^i &= 0,\\
        \partial_t(\rho v^i)+\partial_j\left(\rho v^iv^j+\tau^{ij}\right) & = 0,
        \label{eq:Continuity}
\end{aligned}
\end{equation}
where 
\begin{equation}
\begin{aligned}
    j_i & = qv^i-\rho\partial_i\left(\frac{q}{\rho}\right),\\
    \tau_{ij} & = \rho\delta_{ij}-2\rho_+\partial_{(i}v_{j)}-\left(\rho_+-\rho_-\right)\partial_i\partial_j\log\rho.
    \end{aligned}
    \label{eq:constitutive-relations}
\end{equation}
These are simply continuity equations for the mass, charge and momentum of a compressible fluid up to first order in derivatives together with the addition of a single second order term. The transport coefficients that follow from them are,
\begin{equation}
    \mathcal{P} = \boldsymbol{\mathcal{P}} = \rho,\quad \eta =\boldsymbol{\eta} = \frac{s}{4\pi},\quad \zeta = 0, \quad \kappa_q = \frac{\boldsymbol{\kappa}_q}{n} = \frac{Ts}{4\pi}\left(\frac{Ts}{\rho}\right)^2.
\end{equation}
These are the shear viscosity $\eta$, bulk viscosity $\zeta$ and heat conductivity $\kappa_q$. In terms of our rescaled coordinates we have $c_s=1$, which explains the relation between the pressure and the (rescaled) mass density.

As opposed to a hydrodynamic theory
these equations capture all the physics in the regime where $\mathbf{k}/\mathbf{T}\sim 1/\sqrt{n}$ instead of order by order in a series expansion. In other words, this theory corresponds to a hydrodynamic theory in which all the transport coefficients but a handful of them identically vanish.

Furthermore, we require some notion of an out-of-equilibrium entropy in order to study its time evolution in the collisions. In gravity, a direct candidate for it is simply the apparent horizon area. For the charged case we have that the entropy in \eqref{eq:Thermo} does indeed satisfy the second law,
\begin{equation}
\partial_t s + \partial_i\left(s v^i+\kappa_q\frac{\mu}{T}\partial^i\left(\frac{\mu}{T}\right)\right)\geq 0,
\end{equation}
which is purely originated by the diffusion of charge. Additionally, one can identify a notion of entropy density at first order in $1/n$ which can be written solely in terms of quantities at leading order in $n$ \cite{Andrade:2020ilm},
\begin{equation}
    s_1 = -4\pi\left(\frac{1}{2}\rho v_iv^i+\frac{1}{2\rho}\partial_i\rho\partial^i\rho+\rho\log\rho\right).
    \label{eq:entropy_neutral}
\end{equation}
For in-equilibrium configurations, only the logarithmic term survives and $\rho$ is just a constant. This means that this entropy scales with the volume, as it should. The second law that it satisfies is,
\begin{equation}
    \partial_t s_1 + \partial_i\left(s_1 v^i-4\pi\left(v_j\tau ^{ij}\vert_{q=0}+\partial_j\rho\partial^jv^i\right)\right)\geq 0,
\end{equation}
where $\tau^{ij}\vert_{q=0}$ is the shear stress tensor \eqref{eq:constitutive-relations} with the charge density set to zero. The origin of entropy generation in this case is associated to viscous dissipation. The total entropy is then given by the combination
\begin{equation}
    s_\text{tot} = s+\frac{1}{n}s_1+\mathcal{O}\left(\frac{1}{n^2}\right),
    \label{eq:total_entropy}
\end{equation}
which implies that viscous dissipation is $1/n$ suppressed with respect to charge diffusion. In the neutral case, $s=4\pi\rho$ becomes a constant and all entropy variations come from viscous dissipation, $s_1$.

From now on we will focus on equations \eqref{eq:Large-D-effecttive-equations} as they turn out to be better behaved numerically. In order to obtain a 4+1-dimensional collision \footnote{Notice that we are actually solving a 3+1-dimensional PDE system, because the radial direction has already been integrated when deriving equations \eqref{eq:Large-D-effecttive-equations}.}, we choose to work with nontrivial dependence along three of the $n-1$ horizon directions. The whole set of simulations where done using the code \texttt{Chihuahua} \cite{Chihuahua-2022}, written in the \texttt{Julia} language. It can use both pseudospectral \texttt{FFT} differentiation and finite differences for the spatial derivatives, with a fixed time step \texttt{RK4} time evolution algorithm. In all simulations in this paper, we use single-domain \texttt{FFT} differentiation. We chose $L_x=L_y=L_z=100$, with $N_x=N_y=50$ and $N_z=150$, and the time step is taken to be $\Delta t = 0.1$ in all cases. Running each simulation up to $t_{end} = 30$ took about 20 minutes on a single Intel Core i7-10750H at 2.60GHz CPU, a short time compared to the typical AdS$_5$ shock collision simulations, allowing for a scan over possible different scenarios. The reduced computational cost also allowed us to follow the collision to later times than in previous 4+1-dimensional collisions. Testing for the code is provided in Appendices \ref{app:QNM} and \ref{app:convergence}, using pseudospectral differentiation.

\section{Collisions}
\label{sec:Collisions}

In this section we present the result of colliding two Gaussian blobs of mass that follow the equations of motion \eqref{eq:Large-D-effecttive-equations}. The initial data for the mass, charge and momentum density profiles is given by, 
\begin{equation}
\begin{aligned}
    \rho(0,\vec x) = & 1 + \delta\rho \left\{\exp\left[-\frac{(x-\delta x)^2+y^2}{\sigma_{T1}}-\frac{(z-\delta z)^2}{\sigma_{L1}}\right]+\exp\left[-\frac{(x+\delta x)^2+y^2}{\sigma_{T2}}-\frac{(z+\delta z)^2}{\sigma_{L2}}\right]\right\},\\
    q(0,\vec x) = & q_0+ \delta q \left\{\exp\left[-\frac{(x-\delta x)^2+y^2}{\sigma_{T1}}-\frac{(z-\delta z)^2}{\sigma_{L1}}\right]+\exp\left[-\frac{(x+\delta x)^2+y^2}{\sigma_{T2}}-\frac{(z+\delta z)^2}{\sigma_{L2}}\right]\right\},\\
    p_z(0,\vec x) = & \delta p \left\{-\exp\left[-\frac{(x-\delta x)^2+y^2}{\sigma_{T1}}-\frac{(z-\delta z)^2}{\sigma_{L1}}\right]+\exp\left[-\frac{(x+\delta x)^2+y^2}{\sigma_{T2}}-\frac{(z+\delta z)^2}{\sigma_{L2}}\right]\right\},\\
    p_x(0,\vec x) = &  p_y(0,\vec x) = 0,
\end{aligned}
\end{equation}
where the axes are oriented in such a way that the center of the blobs is contained in the $(x,z)$-plane and the direction of the collision is $z$. $\sigma_L$ is the squared width of the Gaussian in the direction of collision while $\sigma_T$ corresponds to the squared width in the transverse directions.

All collisions were done choosing $\delta \rho = 20$, $\delta p = 60$. The rest of parameters were varied. We set the background charge density as $q_0=0$ in all cases but one, in which we use $q_0 > 0$ in order to allow for charge propagation via the sound mode. We considered several kinds of initial data, as listed in Table \ref{tab:initial_data}.
\begin{table}[thpb]
  \centering
  \begin{tabular}{|c|c|c|c|c|c|c|}
    \hline
    \multicolumn{7}{|c|}{Initial data configurations} \\ \hline \hline
    & $(\sigma_L, \sigma_T)_1$ & $(\sigma_L, \sigma_T)_2$ & $q_0$ & $\delta q$ & $\delta x$ & $\delta z$ \\ \hline \hline
    {\it spherical}          & (10, 10)   & (10, 10)   & 0   & (0, 6, 10) & 0 & 10 \\ \hline
    {\it neutral oblate}     & (10, 50)   & (10, 50)   & 0   & 0          & (0, 2, 5, 8, 10) & 10 \\ \hline
    {\it charged oblate}     & (10, 50)   & (10, 50)   & 0   & 6          & (0, 2, 5, 8, 10) & 10 \\ \hline
    {\it unequal}            & (4, 4)   & (10, 50)   & 0   & 6          & 0 & 10 \\ \hline
    {\it quasi-spherical}    & (30, 50/$\sqrt{3}$) & (30, 50/$\sqrt{3}$) & 0   & 6          & 2 & 15 \\ \hline
    {\it charged background} & (10, 50)   & (10, 50)   & 0.3 & 6          & 2 & 10 \\ \hline
  \end{tabular}
  \caption{Symmary of parameters in inital data configurations. All of them share $\delta\rho=20$, $\delta p = 60$. Through the text we will refer to the different collisions by the names given in this table. \label{tab:initial_data}}
\end{table}
The values of $\sigma_L$ and $\sigma_T$ in the {\it quasi-spherical} blobs are chosen so that the total mass is the same as in the {\it charged oblate} case, which will be relevant when studying entropy growth.

Contrary to what happens at finite $D$, we cannot parametrically suppress the background horizon with respect to the amplitude of the Gaussian blobs. Therefore, the dissipation of the blobs cannot be reduced arbitrarily. Wider Gaussians (at fixed amplitude) lead to longer lived blobs, but require a larger domain, which increases the computational cost. We found that the values of $(\sigma_L,\sigma_T)$ chosen for {\it oblates} in Table \ref{tab:initial_data} are a good compromise.

\subsection{Overview of collisions}

\begin{figure}[thpb]

\centerline{\includegraphics[width=\textwidth]{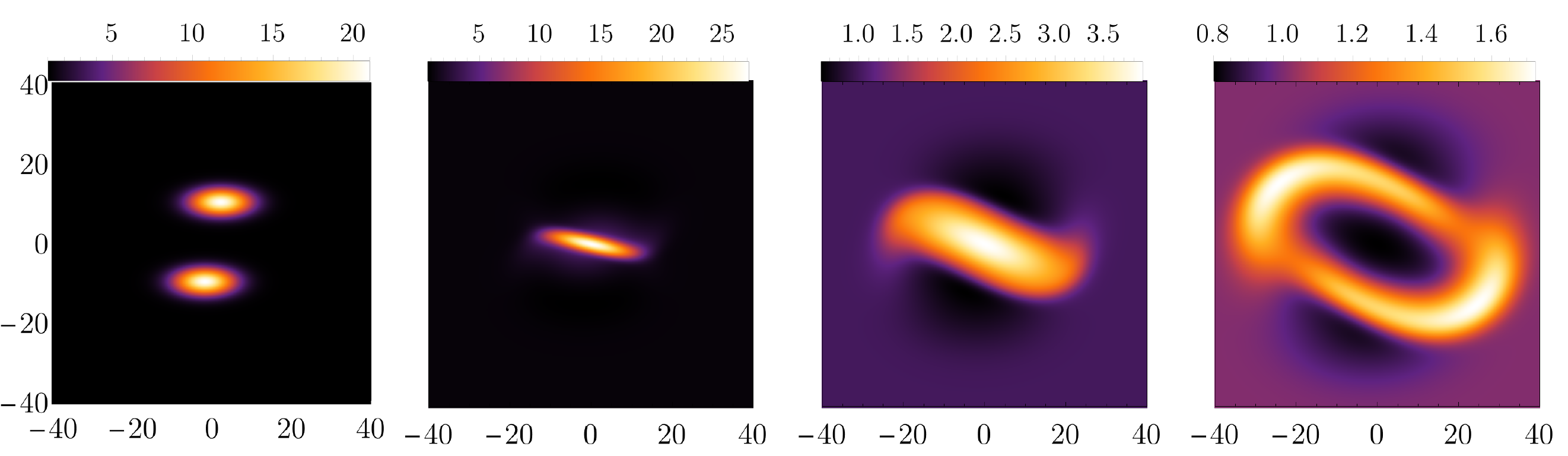}
\put(-435,55){\mbox{{$z$}}}
\put(-376,-7){\mbox{{$x$}}}
\put(-271,-7){\mbox{{$x$}}}
\put(-163,-7){\mbox{{$x$}}}
\put(-55,-7){\mbox{{$x$}}}
}

\centerline{\includegraphics[width=\textwidth]{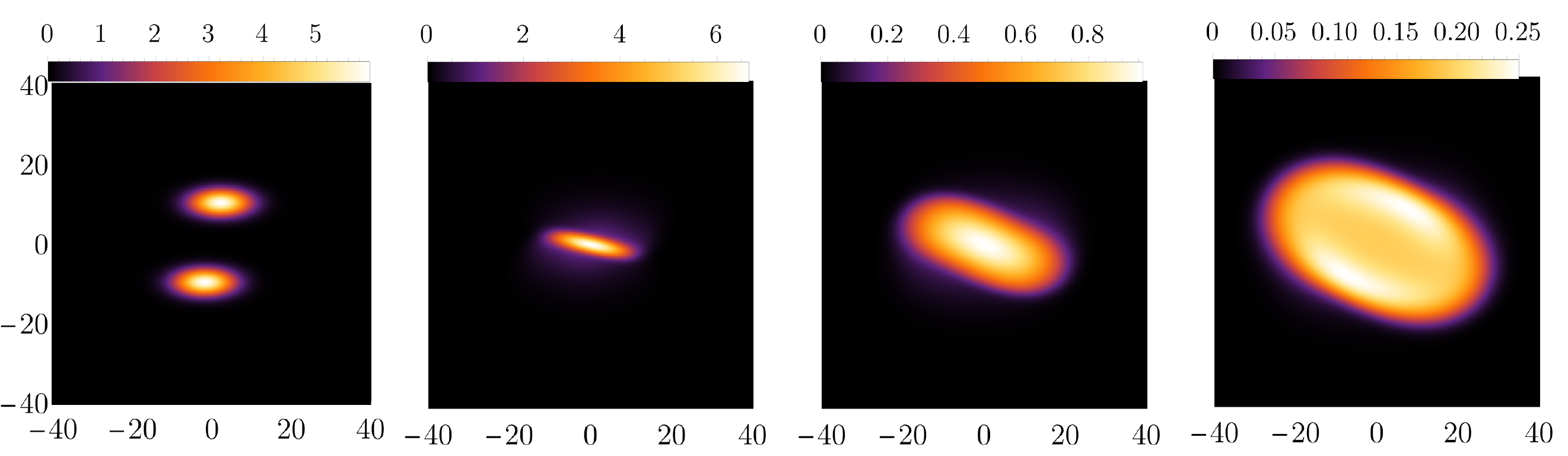}
\put(-435,55){\mbox{{$z$}}}
\put(-376,-7){\mbox{{$x$}}}
\put(-271,-7){\mbox{{$x$}}}
\put(-163,-7){\mbox{{$x$}}}
\put(-55,-7){\mbox{{$x$}}}
}

\centerline{\includegraphics[width=\textwidth]{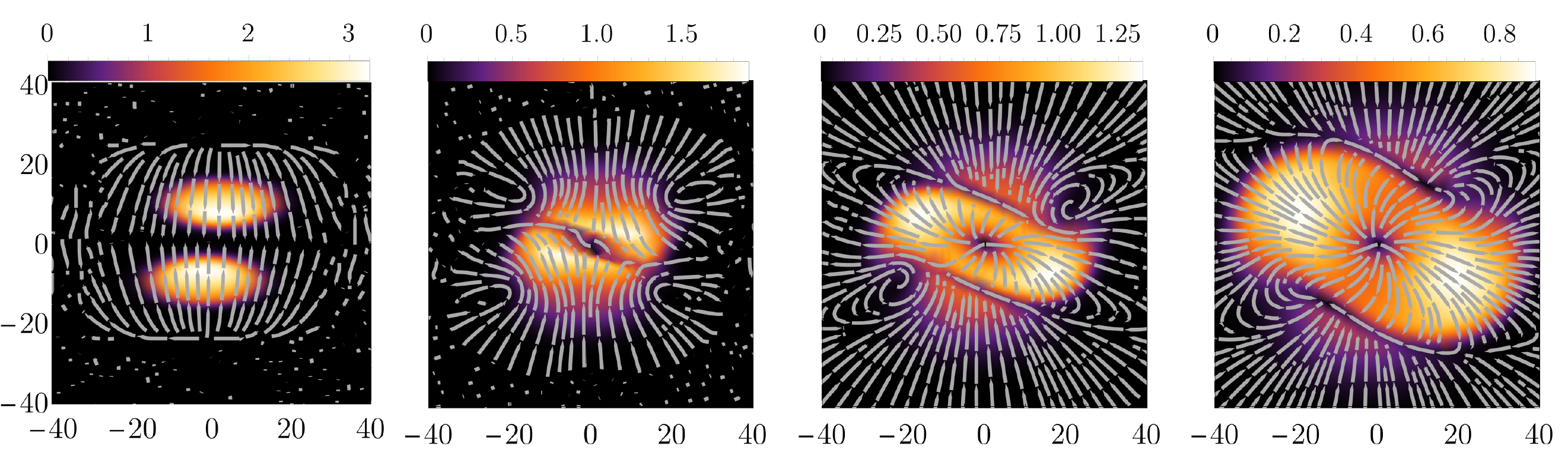}
\put(-435,55){\mbox{{$z$}}}
\put(-376,-7){\mbox{{$x$}}}
\put(-271,-7){\mbox{{$x$}}}
\put(-163,-7){\mbox{{$x$}}}
\put(-55,-7){\mbox{{$x$}}}
}

\caption{Mass density (top row), charge density (middle row) and fluid velocity (bottom row) at the collision plane for a 4+1-dimensional {\it charged oblate} blob collision with $\delta x = 2$ from Table \ref{tab:initial_data}. The snapshots correspond to $t = (0,4,9,15)$.}
\label{fig:snapshots_asymmetric}
\end{figure}

In Figure~\ref{fig:snapshots_asymmetric} one can find snapshots corresponding to the collision of {\it charged oblate} blobs with $\delta x = 2$. Initially, the blobs approach each other until they collide at $t = t_{c} \approx 3.11$ to form a blob of mass that is highly compressed in the direction of the collision, $z$. We define the collision time, $t_c$, as the time in which the mass density reaches is maximum value. After the collision, the resulting blob of mass expands and the mass density at the collision site decreases. The shape of the expanding blob is far from symmetric, as shown in Figure \ref{fig:snapshots_asymmetric}, the mass density is flowing outwards in a roughly elliptical shape. The minor axis of such ellipse coincides with the line joining the initial blobs. Up to $t \approx 10$, both the mass density and the charge profile follow similar evolution patterns. However, at later times the charge density will simply follow a diffusion pattern, while the mass density will continue to propagate away from the collision site leaving a depleted region in its middle. The reason behind such a difference is that the background metric is neutral, so charge density propagation modes cannot be excited and only diffusion takes place. On the other hand, the mass density diffuses but it also propagates\footnote{These notions apply strictly in the linear regime of small perturbations, but the nonlinear physics involved in the collisions seems to retain some of these features.}. After $t_c$, a high dissipation stage takes place. While the mass and charge density have barely decreased until $t\approx 4$, they both fall down by an order of magnitude by $t \approx 9$.

\begin{figure}[thpb]
\centerline{\includegraphics[width=\textwidth]{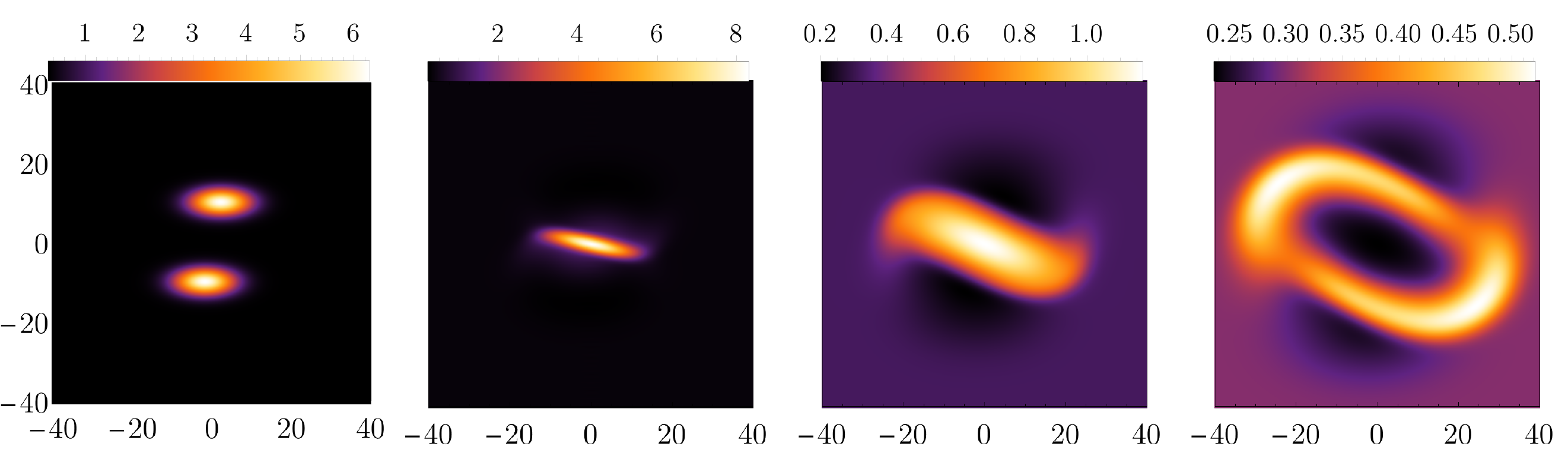}
\put(-435,55){\mbox{{$z$}}}
\put(-376,-7){\mbox{{$x$}}}
\put(-271,-7){\mbox{{$x$}}}
\put(-163,-7){\mbox{{$x$}}}
\put(-55,-7){\mbox{{$x$}}}
}

\caption{Charge density at the collision plane at $t=(0,4,9,15)$ for the {\it charged background} collision from Table \ref{tab:initial_data}.}
\label{fig:snapshots_bckg_charge}
\end{figure}

Figure \ref{fig:snapshots_bckg_charge} shows the charge density in a {\it charged background} collision. In this case, charge can both diffuse and propagate on the background horizon, and so its evolution very much resembles that of the mass density in Figure \ref{fig:snapshots_asymmetric}. We see that the charge blob resulting from the collision ends up fragmenting in a similar fashion to the mass density. 

\begin{figure}[thpb]
\centerline{\includegraphics[width=\textwidth]{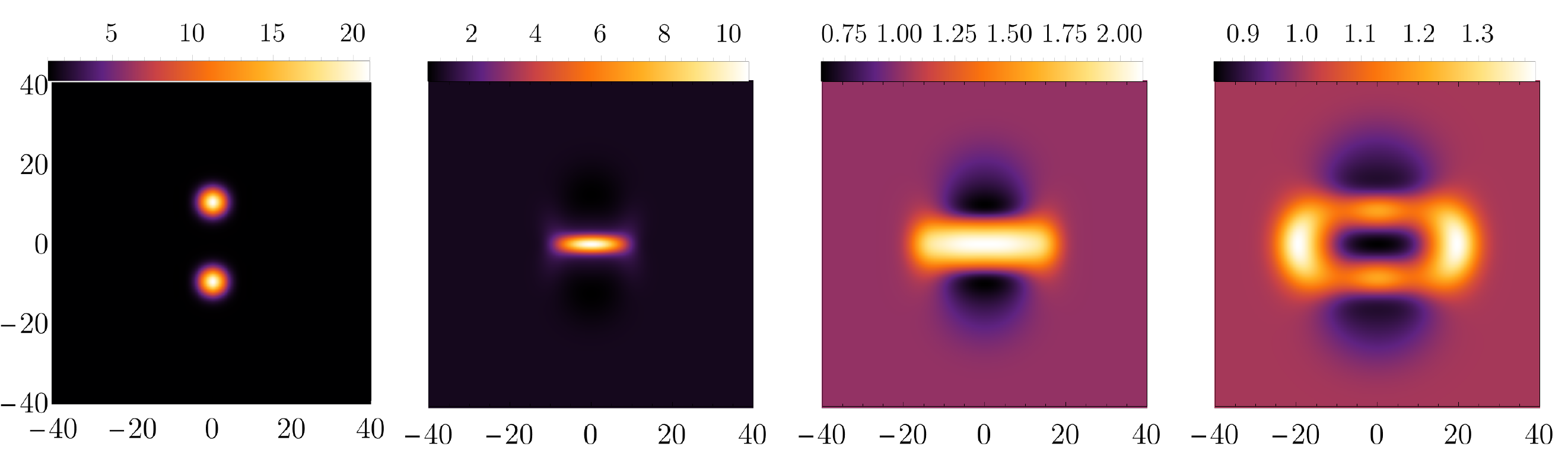}
\put(-435,55){\mbox{{$z$}}}
\put(-376,-7){\mbox{{$x$}}}
\put(-271,-7){\mbox{{$x$}}}
\put(-163,-7){\mbox{{$x$}}}
\put(-55,-7){\mbox{{$x$}}}
}

\caption{Mass density at the collision plane at $t=(0,4,8,12)$. The collision corresponds to the head-on {\it spherical} blob collision from Table \ref{tab:initial_data}.}
\label{fig:snapshots_symmetric}
\end{figure}

In Figure \ref{fig:snapshots_symmetric} one can find a head-on ($\delta x = 0$) collision of {\it spherical} blobs. As one might have expected, dissipation of the blobs during the early stage in which they approach each other is higher in this case, due to the smaller size of the blobs. In fact, by $t \approx 4$, the mass density has dropped to half of its original value. As in the {\it charged oblate} case, the blob resulting from the collision is elliptical in shape, which ends up fragmenting into two lumps of mass that travel along the major axis of the ellipse.

\begin{figure}[thpb]
\centerline{\includegraphics[width=\textwidth]{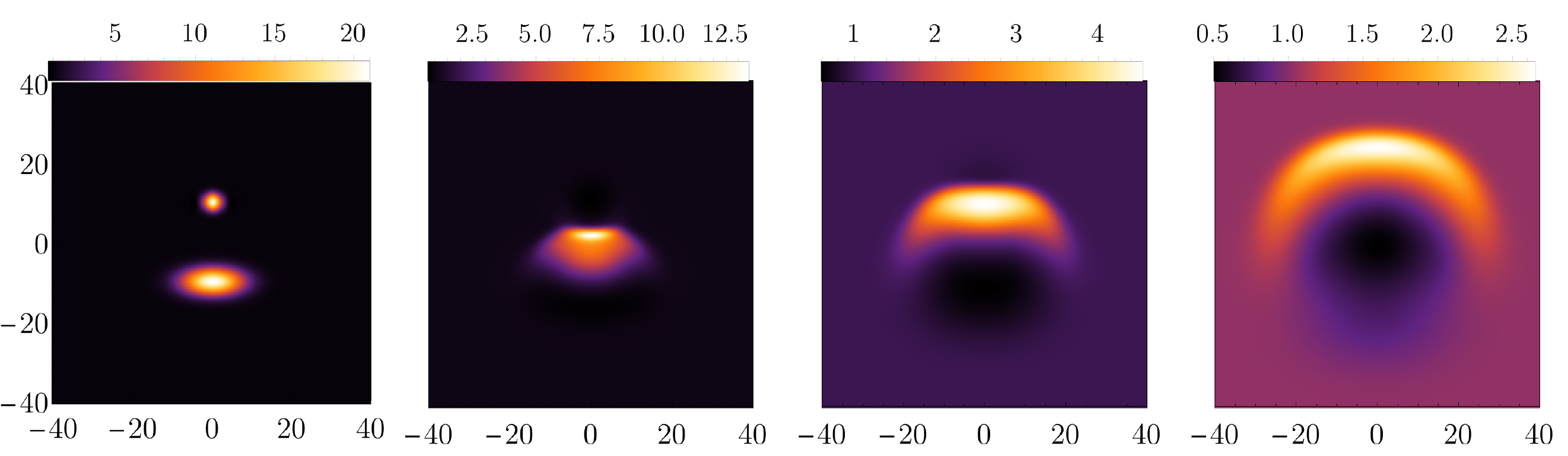}
\put(-435,55){\mbox{{$z$}}}
\put(-376,-7){\mbox{{$x$}}}
\put(-271,-7){\mbox{{$x$}}}
\put(-163,-7){\mbox{{$x$}}}
\put(-55,-7){\mbox{{$x$}}}
}

\caption{Mass density at the collision plane at $t=(0,3,7,13)$. The collision corresponds to the {\it unequal} blob collision from Table \ref{tab:initial_data}.}
\label{fig:snapshots_unequal}
\end{figure}

Finally, Figure \ref{fig:snapshots_unequal} shows the snapshots of {\it unequal} blobs collision. The result of such a collision is completely different to the previous ones. Its appearance is similar to the propagation of a single blob, perturbed by the collision with a smaller blob. As time runs, the blob diffuses leaving the usual depletion of mass behind it and the shape of the front relaxes to the shape that a freely moving blob would have.
\subsection{Impact parameter dependence}

\begin{figure}[thpb]
\centerline{
\includegraphics[width=0.5\textwidth]{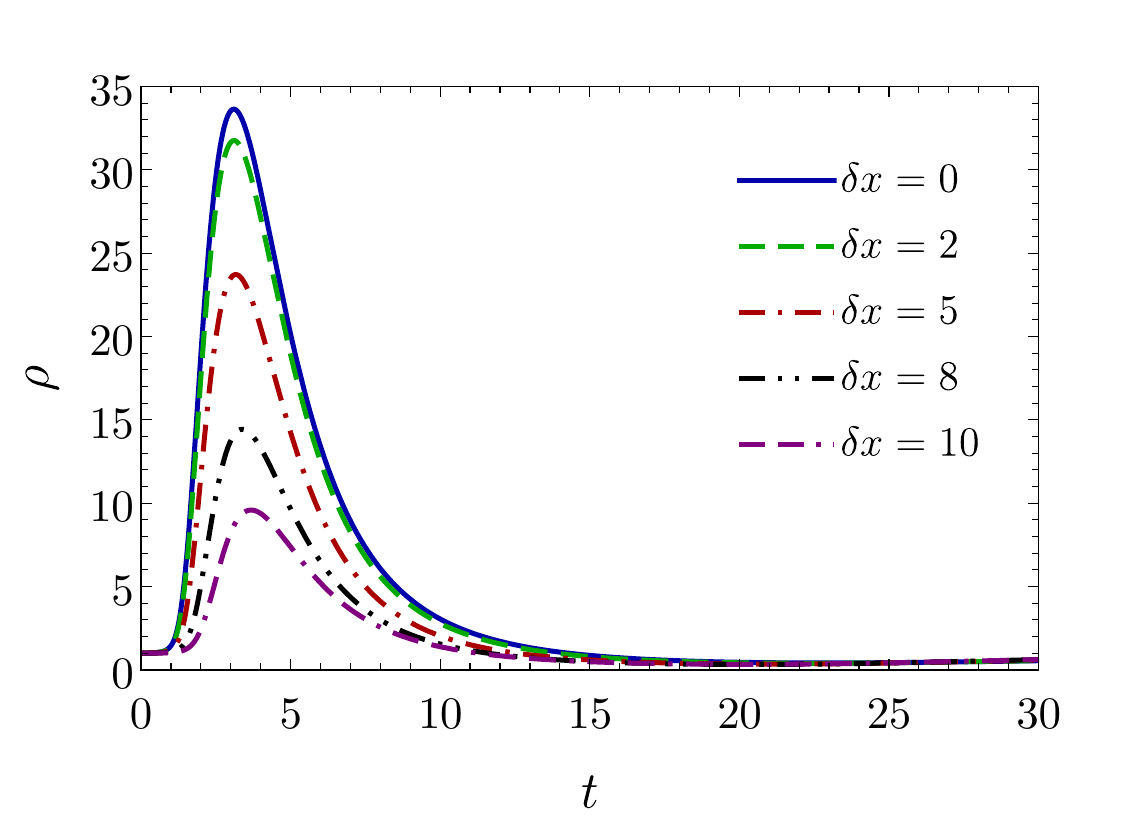}
\includegraphics[width=0.5\textwidth]{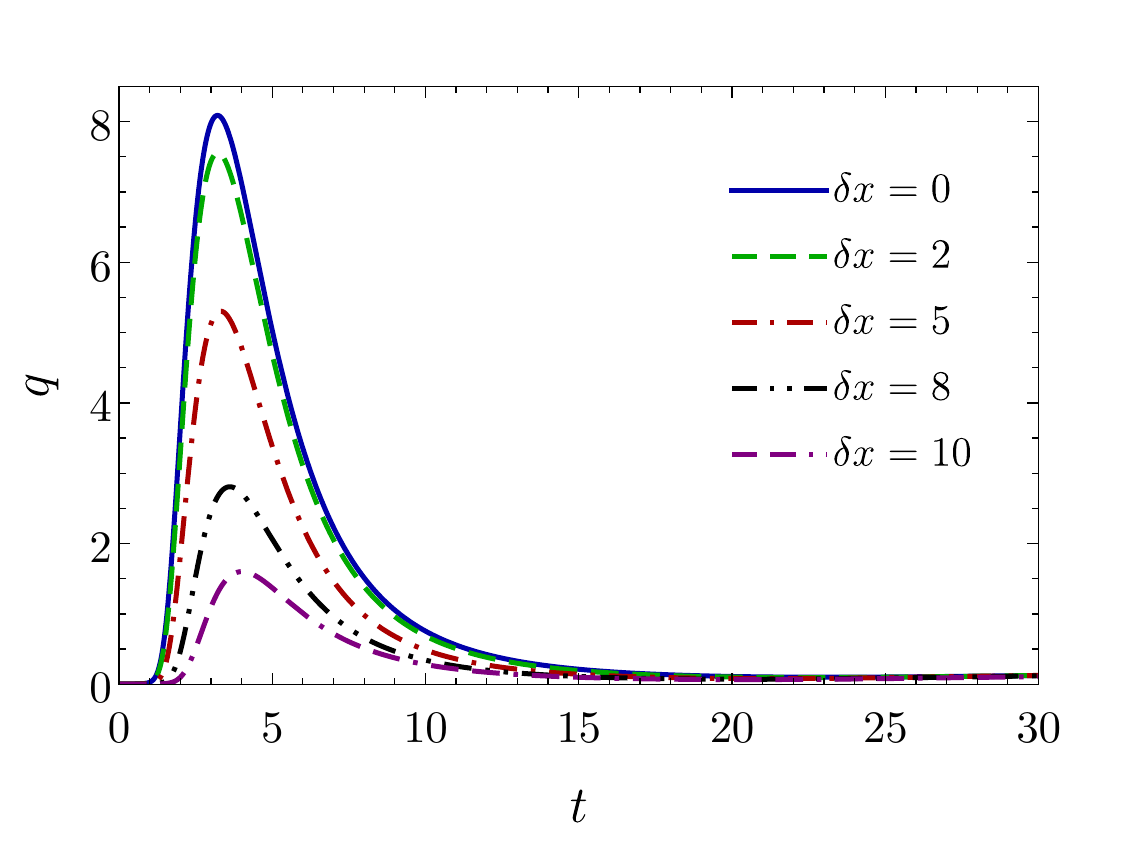}
}
\centerline{
\includegraphics[width=0.5\textwidth]{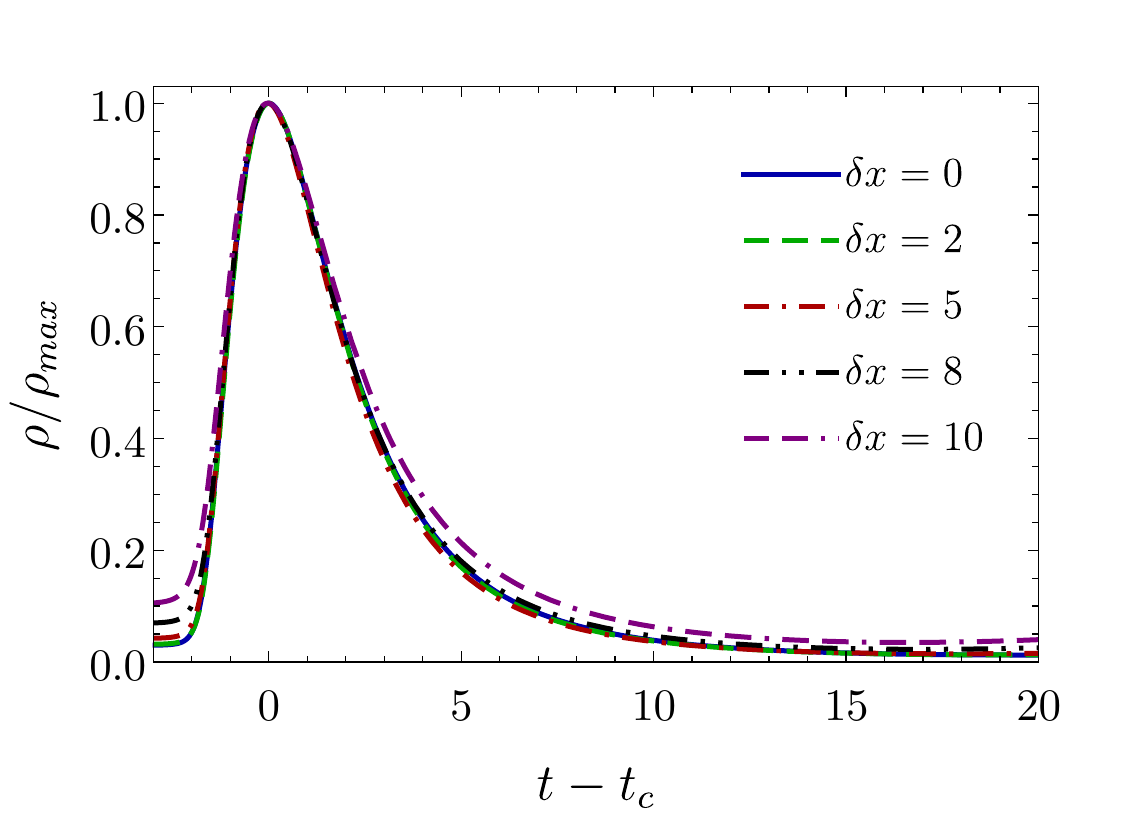}
\includegraphics[width=0.5\textwidth]{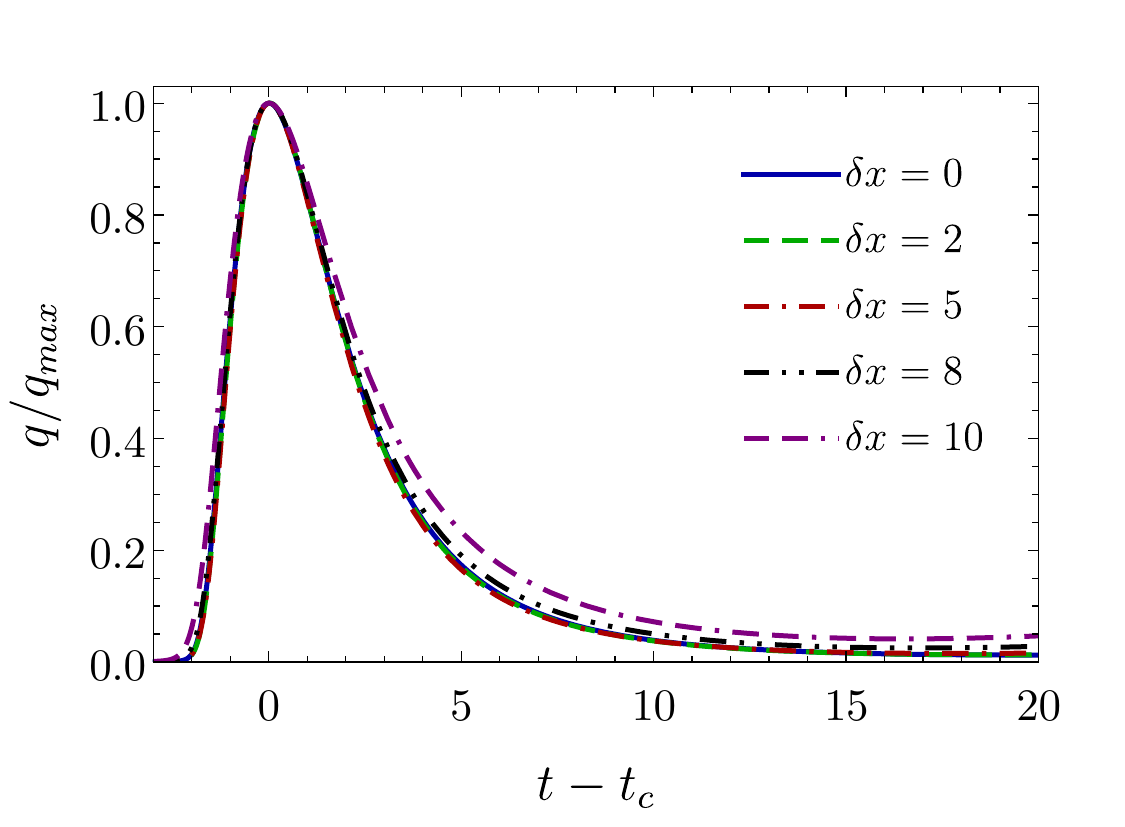}
}
\caption{Top: Mass and charge densities at the origin as a function of time for \textit{charged oblate} collisions with different impact parameters. Bottom: Mass and charge densities normalized by their the maximum, as a function of the elapsed time since the collision.}
\label{fig:dx_dependence}
\end{figure}

In order to better assess the differences that arise from different impact parameters, in Figure \ref{fig:dx_dependence} we show the result of all {\it charged oblate} collisions. In the top panels we observe that, as $\delta x$ grows, the maximum mass and charge densities reached during the collision decrease. This is to be expected, since a bigger separation in the $x$-axis causes the effective overlap between the blobs to be smaller. In other words, the collision is less violent as the impact parameter grows. The collision time $t_c$ is almost insensitive to the value of $\delta x$.

It is instructive to compare the mass and charge profiles at the origin, normalized by their maximum, as a function of time. The result is shown in the bottom panel of Figure \ref{fig:dx_dependence}. All the curves with $\delta x\leq 8$ are very close to each other, while some major discrepancies are observed for the largest impact parameter: $\delta x = 10$. Therefore, collisions whose impact parameter does not surpass the blob transverse width ($\sqrt{\sigma_T}$) can be approximately seen as rescaled versions of one another.

\subsection{Isotropization and hydrodynamization}

Due to the large flow of mass in the direction of collision, large anisotropies are present. As the product of the collision approaches equilibrium, two events take place during the evolution: isotropization and hydrodynamization. The former is simply a consequence of the fact that equilibrium states are isotropic. The latter comes from the expectation that interacting systems whose departure from equilibrium is small can be well characterized by hydrodynamics.

An easy way of testing to what extent a system has isotropized and hydrodynamized is by looking into the evolution of the three different pressures in the system. By three pressures we mean the diagonal components, $\mathcal{P}_x=\tau_{xx}$, $\mathcal{P}_y=\tau_{yy}$ and $\mathcal{P}_z=\tau_{zz}$, of the stress tensor in \eqref{eq:constitutive-relations}. Discrepancies among these pressures are a sign of anisotropy. Also, once the system has hydrodynamized, the hydrodynamic constitutive relations should be a good approximation of $\tau_{ii}$. The way in which we will define hydrodynamization is by comparing the full pressures in the system ($\mathcal{P}_i$) to the viscous hydrodynamic ones ($\mathcal{P}_i^V$). We define $\mathcal{P}_i^V$ as the expressions for $\tau_{ii}$ in \eqref{eq:constitutive-relations} without the inclusion of second order derivative terms,
\begin{equation}
\begin{aligned}
    \mathcal{P}_i & = \rho - 2 \rho_+ \, \partial_i v_i - \left(\rho_+ - \rho_-\right)\partial_i^2 \log\rho,\\
    \mathcal{P}_i^V & = \rho - 2\rho_+ \, \partial_i v_i.
    \end{aligned}
    \label{eq:pressures}
\end{equation}
The large $D$ effective theory can be seen as a hydrodynamic theory up to second order in gradients. Hence, the theory is always in a hydrodynamic regime, but our hydrodynamization time is then a measure of the required time for second order gradients to become negligible. Let us emphasize that we are not performing the time evolution of viscous hydrodynamics. We are instead using the data of the full solutions to evaluate $\mathcal{P}_i^V$, and then comparing it point-wise to $\mathcal{P}_i$. The charge current has no second order derivative terms, so it is always well captured by first-order hydrodynamics.
\begin{figure}[thpb]
\centerline{
\includegraphics[width=0.5\textwidth]{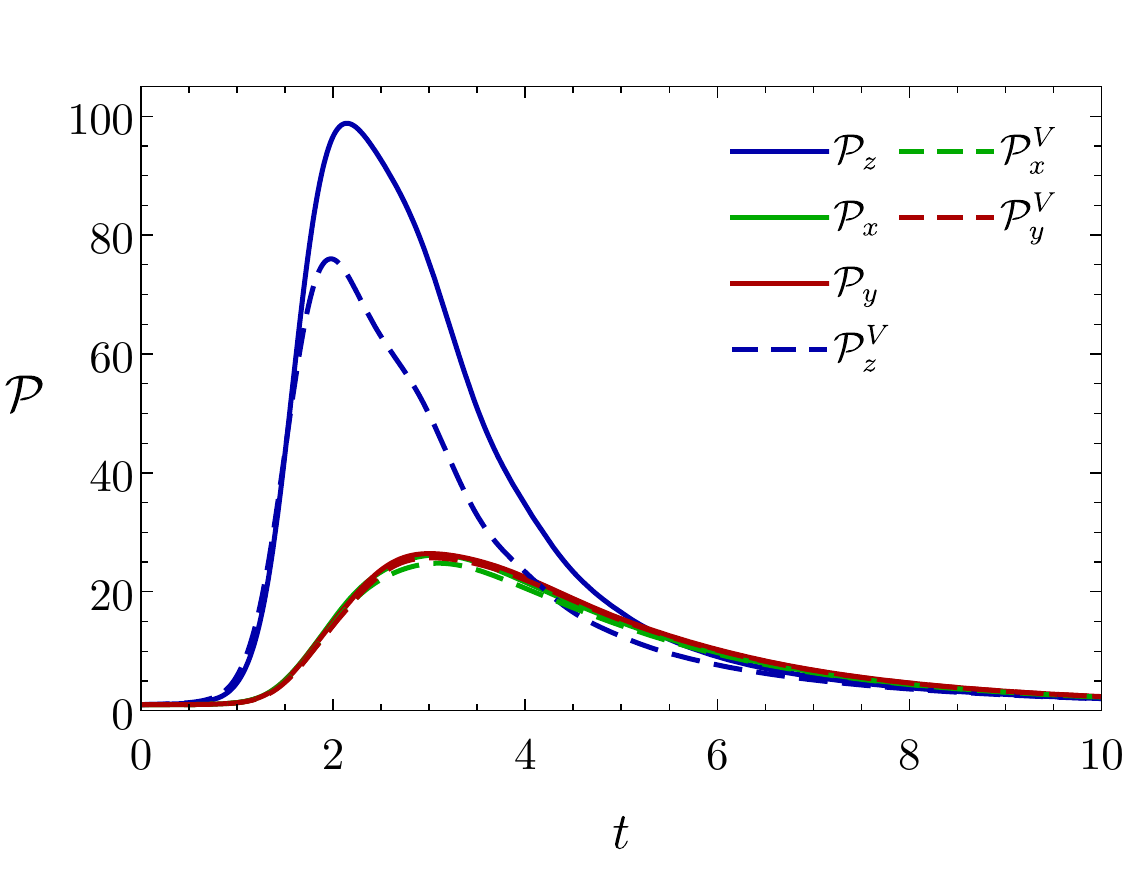}
\includegraphics[width=0.5\textwidth]{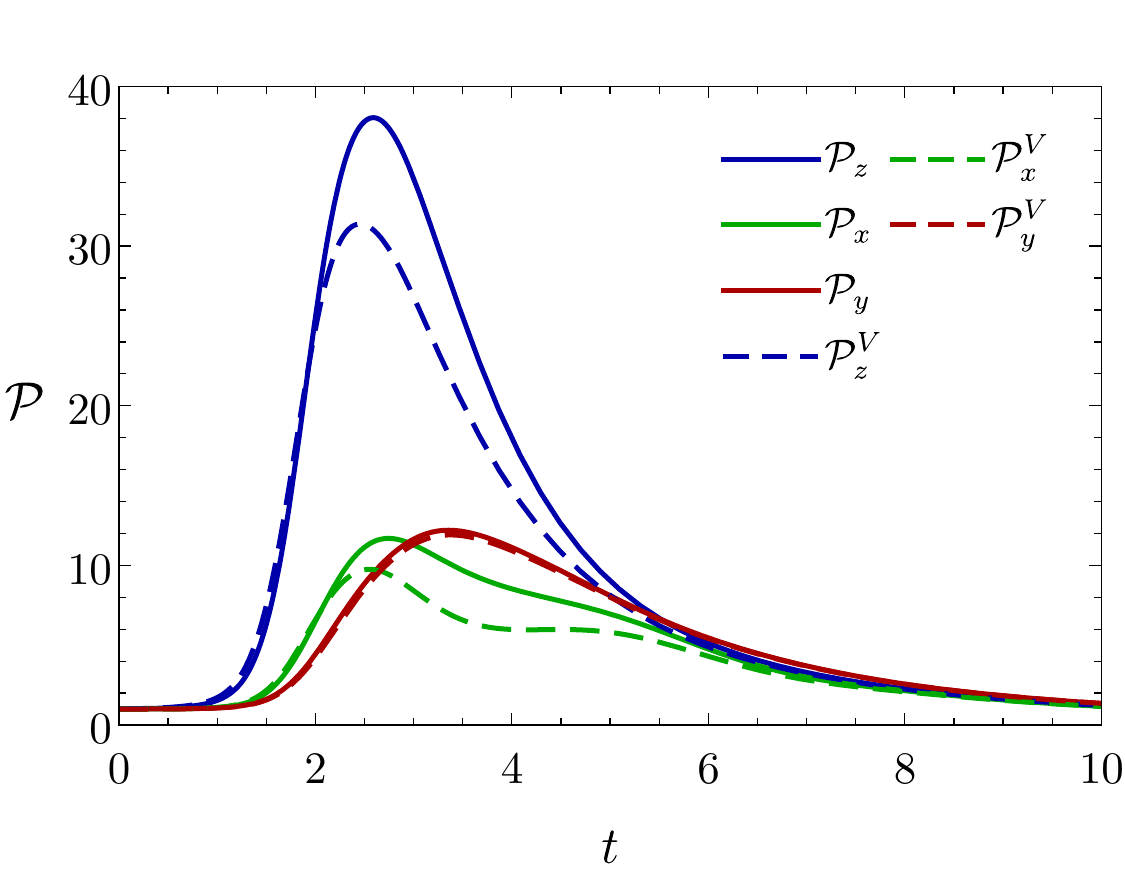}
}
\centerline{
\includegraphics[width=0.5\textwidth]{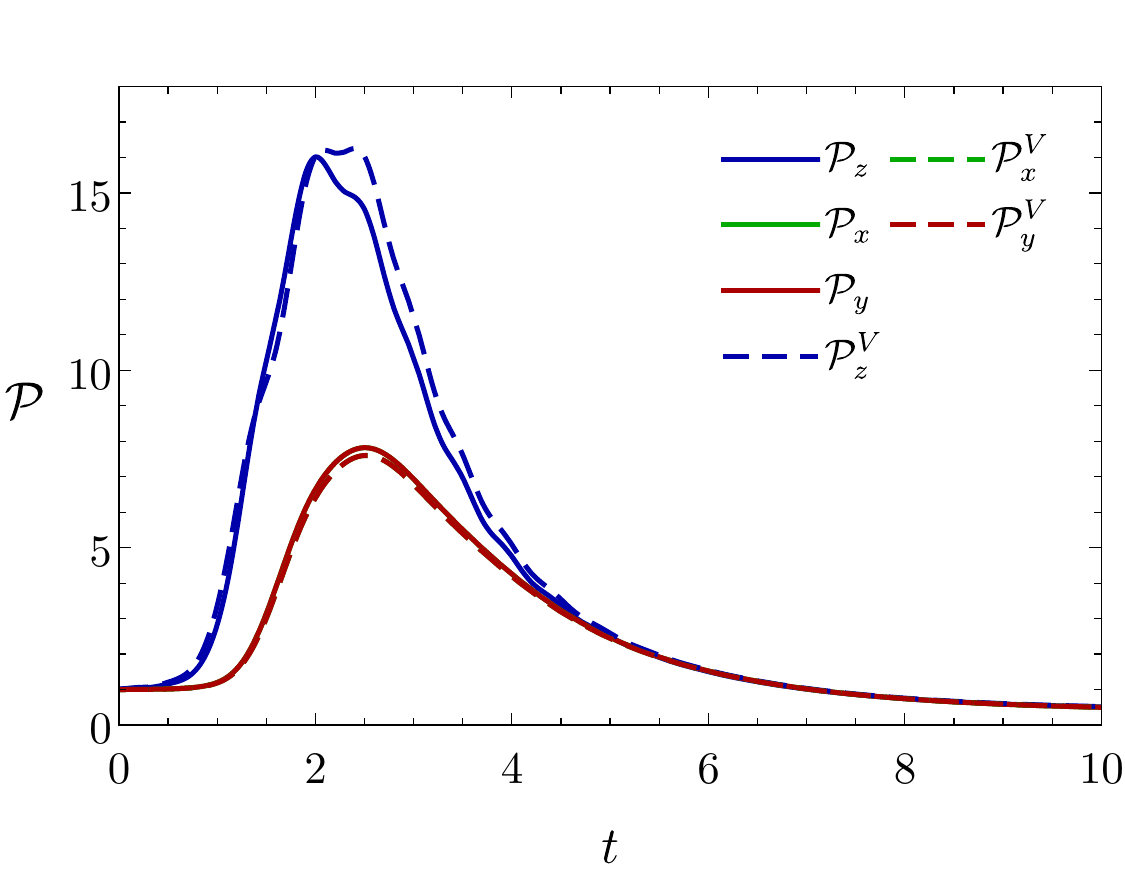}
}
\caption{Pressures along the three active directions as a function of time at the spatial origin. The solid lines represent the full pressures ($\mathcal{P}_i$), while the dashed lines show by the first order hydrodynamic pressures ($\mathcal{P}_i^V$). Top: \textit{Charged oblate} blob collision with impact parameter $\delta x = (2,8)$ for the right and left panels respectively. Bottom: \textit{Unequal} blob collision.}
\label{fig:pressures}
\end{figure}

In Figure \ref{fig:pressures} we show the pressures as a function of time at the collision spot, for {\it charged oblate} blob collisions (with $\delta x = 2, 8$) in the top panels, and for an {\it unequal} blob collision in the bottom one. Continuous lines refer to $\mathcal{P}_i$, while the dashed lines correspond to the viscous hydrodynamic pressures, $\mathcal{P}_i^V$. We focus on the collision point, where largest gradients are expected. 

The top panel of Figure \ref{fig:pressures} illustrates the large anisotropies reached during the collision, with a ratio of longitudinal to transverse pressures that can get slightly over 3. Also, viscous hydrodynamics fails to describe the system at times around $t_c$, especially along the axis of collision. For lower impact parameter collisions, a greater anysotropy is produced and viscous hydrodynamics further departs from the actual value for $\mathcal{P}_i$. Furthermore, the bigger $\delta x$ gets, the more the $x \longleftrightarrow y$ symmetry is broken, so the difference between $\mathcal{P}_x$ and $\mathcal{P}_y$ gets accentuated. For $\delta x$ larger than the blob width, the anysotropy in the $x-y$ plane will presumably decrease.

The results suggest that, after $t \approx 7$, the system becomes very isotropic and well described by viscous hydrodynamics, roughly at the same time. This result differs from the findings in finite $D$ collisions, where hydrodynamization takes place earlier than isotropization \cite{Chesler:2010bi,Casalderrey-Solana:2013aba, Chesler:2015wra}. Since the conclusions extend to the rest of collisions in Table \ref{tab:initial_data}, we relate the discrepancy we found to the large $D$ limit and not to the addition of charge. 

The results for an {\it unequal} blob collision, as shown in the bottom panel of Figure \ref{fig:pressures}, differ from the rest. Throughout the whole process, viscous hydrodynamics approximates better the physics, and it gives an overestimation of the pressure (in opposition to the rest of collisions). The level of anysotropy that is reached is smaller than for equal shock collisions.

\subsection{Charge influence}

\begin{figure}[thpb]
\centerline{
\includegraphics[width=0.5\textwidth]{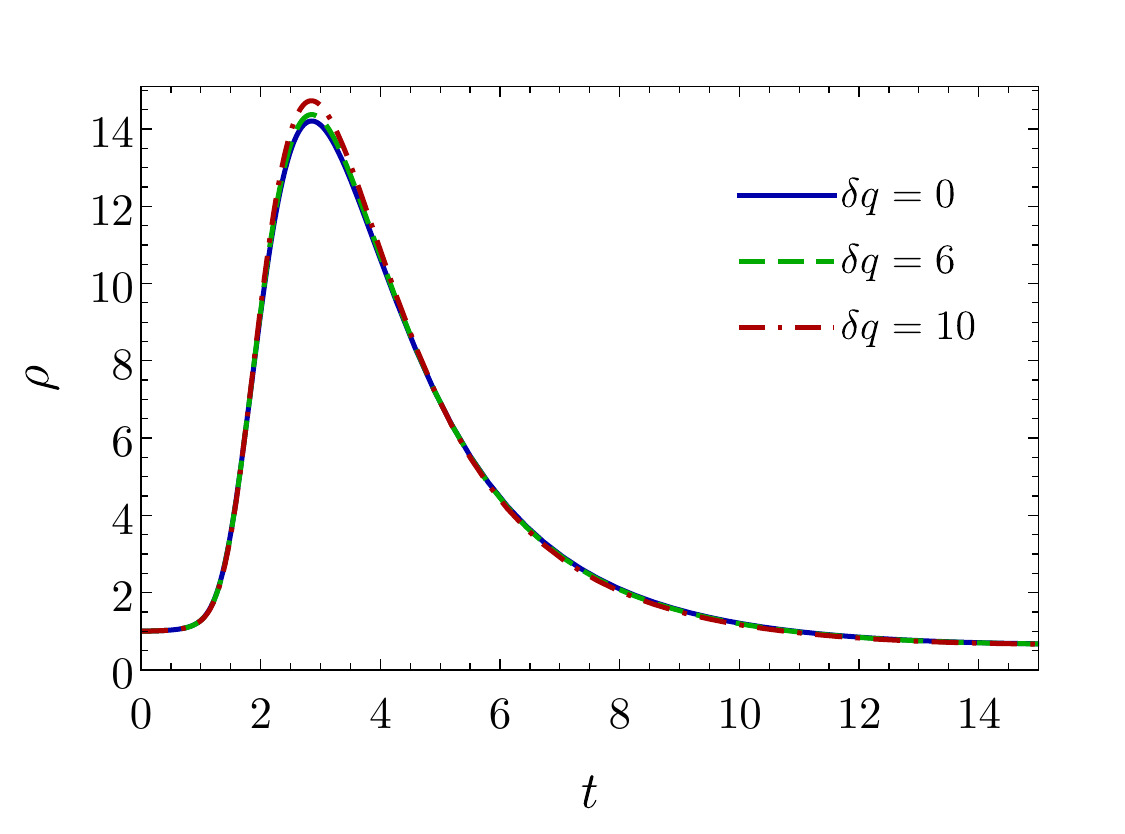}
\includegraphics[width=0.5\textwidth]{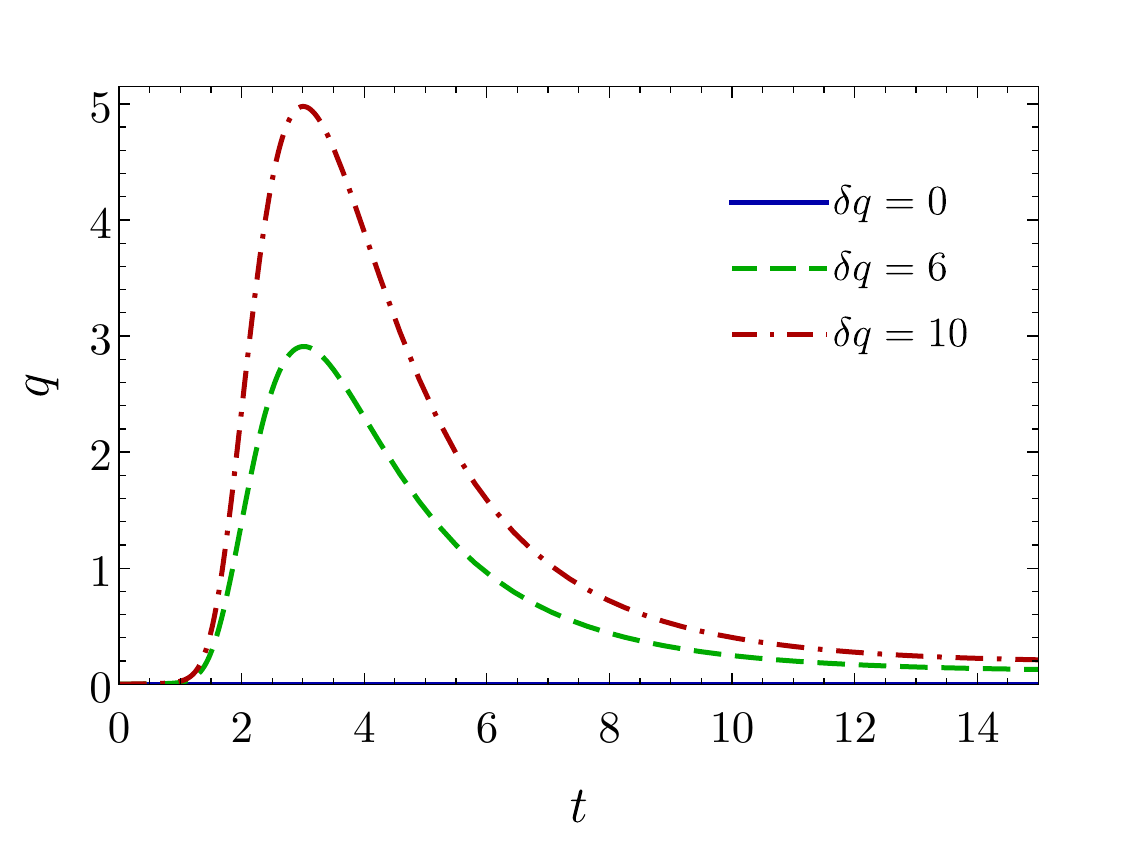}
}
\centerline{
\includegraphics[width=0.5\textwidth]{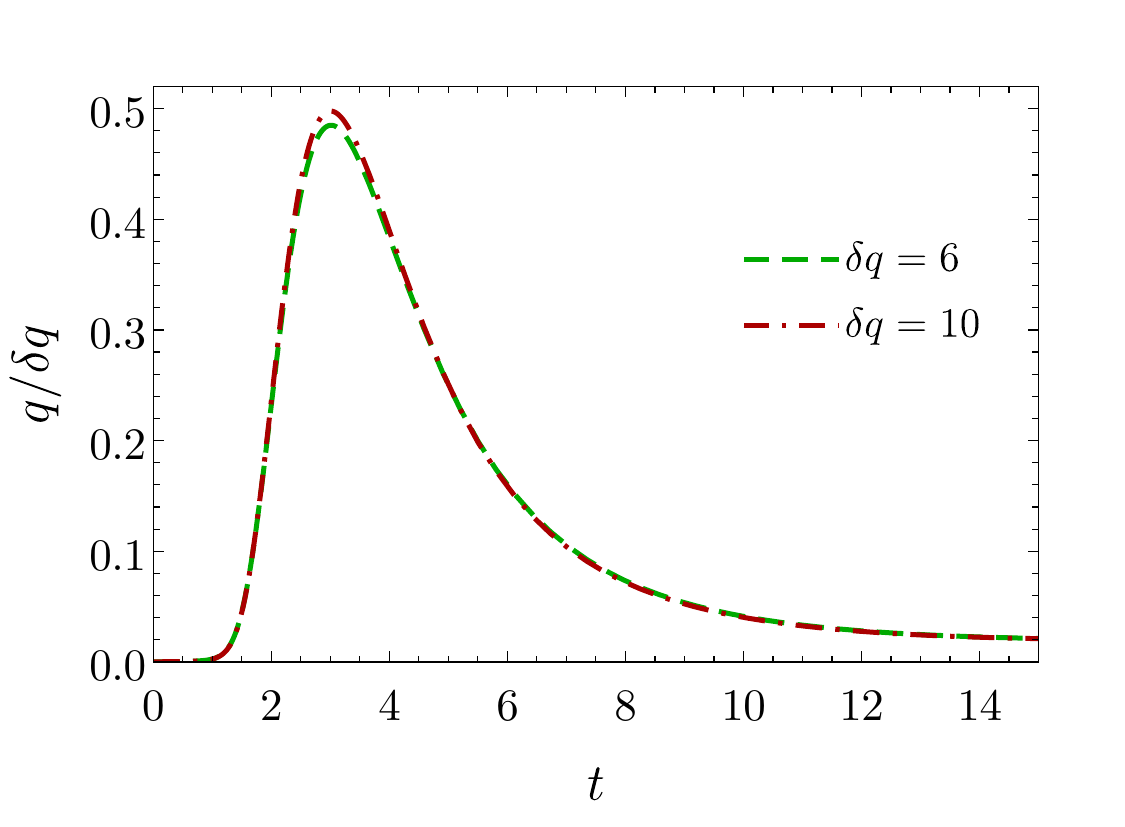}
\includegraphics[width=0.5\textwidth]{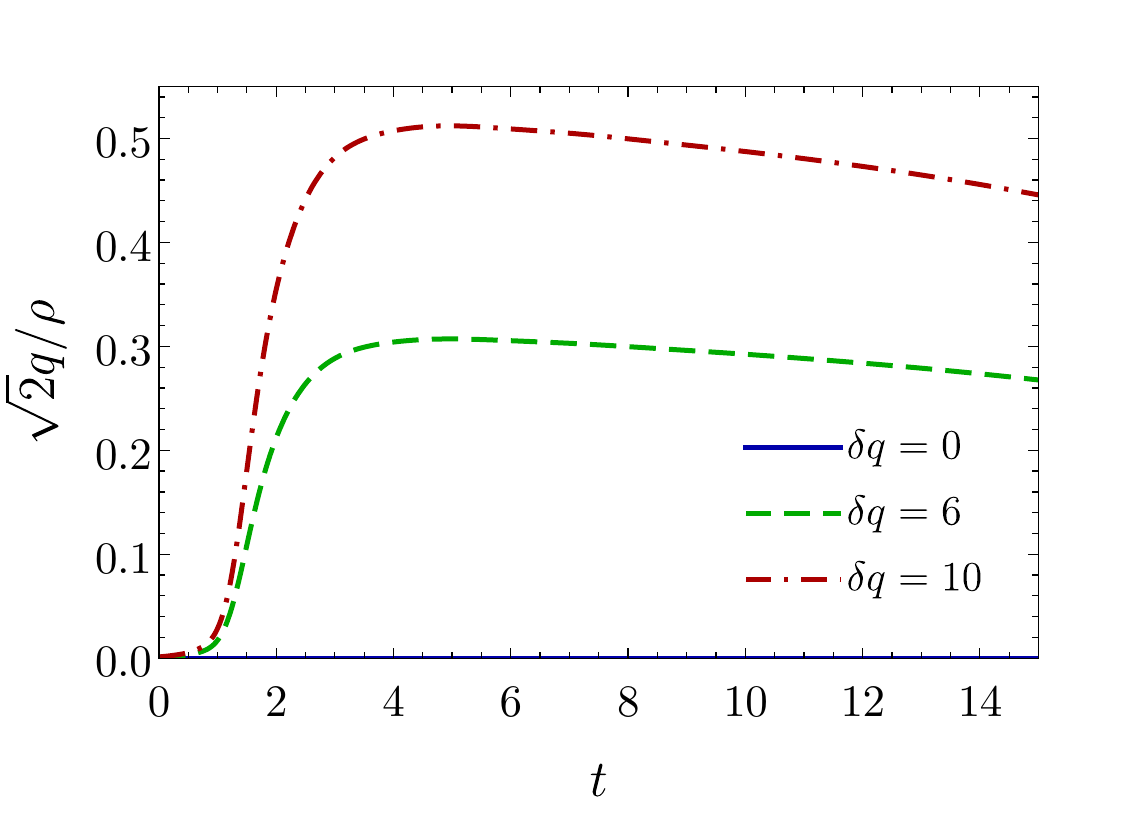}
}
\caption{Mass, charge density, charge density normalized by $\delta q$ and the ratio $\sqrt{2}q/\rho$ at the spatial origin for the head-on \textit{spherical} blob collisions of Table \ref{tab:initial_data}.}
\label{fig:charge_influence_mass_charge_extr}
\end{figure}

In this subsection we compare the results of collisions with different values for the charge of the blobs to study its influence on the dynamics. In Figure \ref{fig:charge_influence_mass_charge_extr} we display the mass, charge density, charge density normalized by the blob charge $\delta q$ and the ratio $\sqrt{2}q/\rho$ at the spatial origin for the {\it spherical} blob collisions, with $\delta q = (0, 6, 10)$.
The behavior of the charge density is qualitatively similar, approximately proportional to $\delta q$, see Figure \ref{fig:charge_influence_mass_charge_extr} bottom-left. Surprisingly, even though the maximum charge density is around a third of the maximum mass density ($\sqrt{2}q/\rho$ gets up to $1/2$) the effect of charge on the mass density is small. The biggest difference takes place at the collision time ($t_c$), where larger values of  the mass density are achieved for larger charge. For $\Delta t\approx 2$ after the collision time, the mass density follows the same evolution as in the neutral collision. The time it takes to the mass density to follow the neutral collision profile is shorter than the isotropization and hydrodynamization timescales. 
\begin{figure}[thpb]
\centerline{
\includegraphics[width=0.5\textwidth]{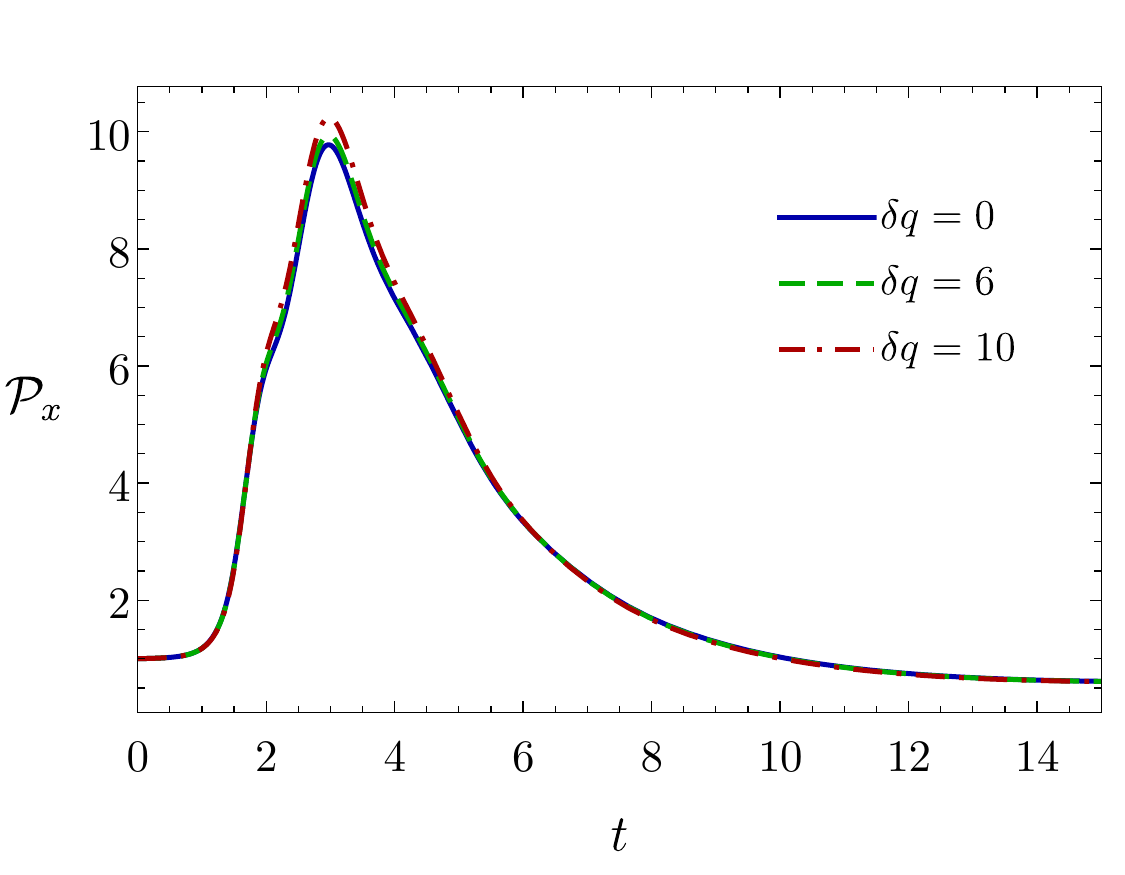}
\includegraphics[width=0.5\textwidth]{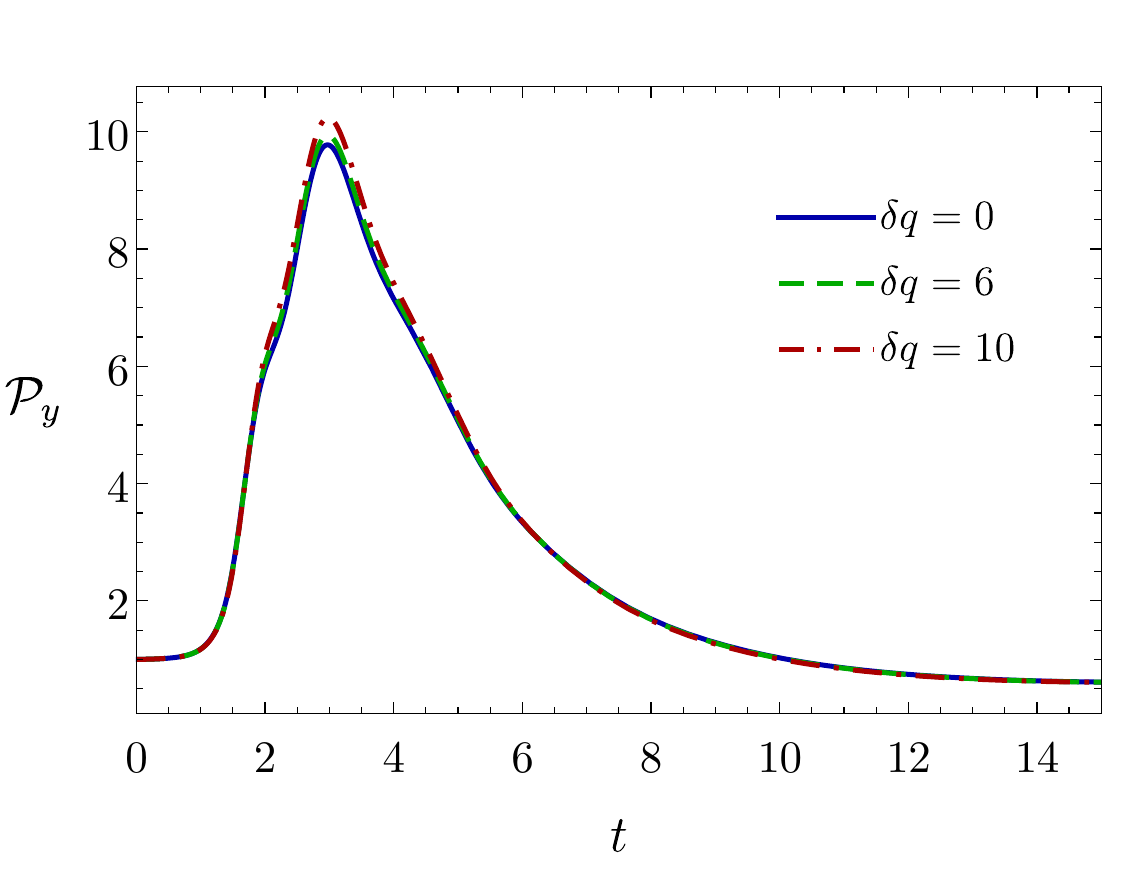}
}
\centerline{
\includegraphics[width=0.5\textwidth]{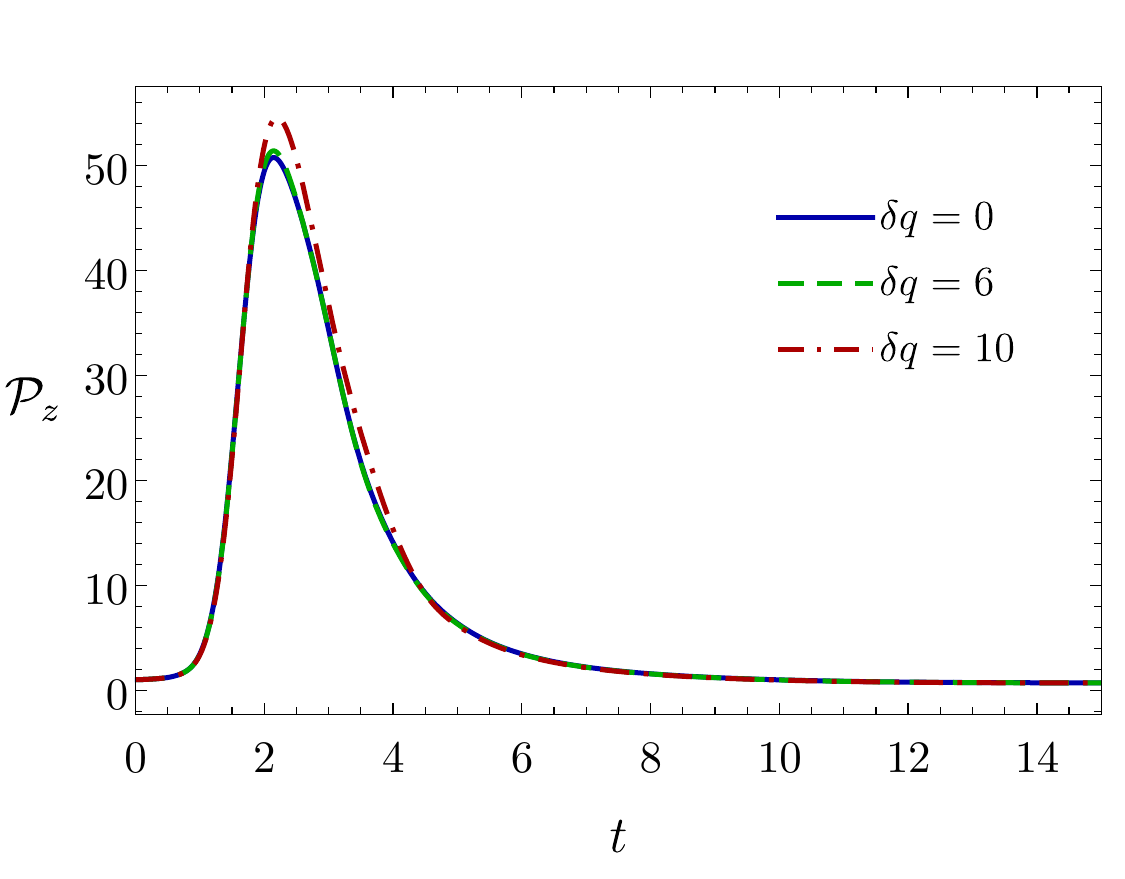}
}
\caption{Pressures along the three active directions ($\tau_{ii}$) at the spatial origin for head-on \textit{spherical} blob collisions from Table \ref{tab:initial_data}.}
\label{fig:charge_influence_pressures}
\end{figure}

Figure \ref{fig:charge_influence_pressures} shows the time evolution of the pressure along all three directions at the collision location. The maximum pressure achieved does also increase with the blob charge $\delta q$, and the pressures follow very closely the result of the neutral collision soon after $t_c$, although $\mathcal{P}_z$ is more sensitive to the presence of charge than any of the previously studied quantities.

Our results therefore suggest that charge does not greatly affect the collision dynamics. Similarly, \cite{Casalderrey-Solana:2016xfq} found that charge does not greatly affect other observables in planar shockwave collisions at finite $D$. Due to technical difficulties, we could not go beyond the maximum charge value presented here, so it is still unknown to us if our conclusions would change for larger $\delta q$. 

\section{Entropy growth}
\label{sec:Entropy}

We now present the details of entropy evolution during collisions. As mentioned earlier, the leading entropy production comes from charge diffusion, while viscous dissipation is $1/n$ suppressed. Therefore, for charged collisions we will focus on the entropy defined in \eqref{eq:Thermo}, while for neutral collisions we take the entropy in \eqref{eq:entropy_neutral}. The results that we present here correspond to the {\it charged oblate} and {\it neutral oblate} collisions in Table \ref{tab:initial_data}.

\subsection{Charged collisions}

\begin{figure}[thpb]
\centerline{
\includegraphics[width=0.8\textwidth]{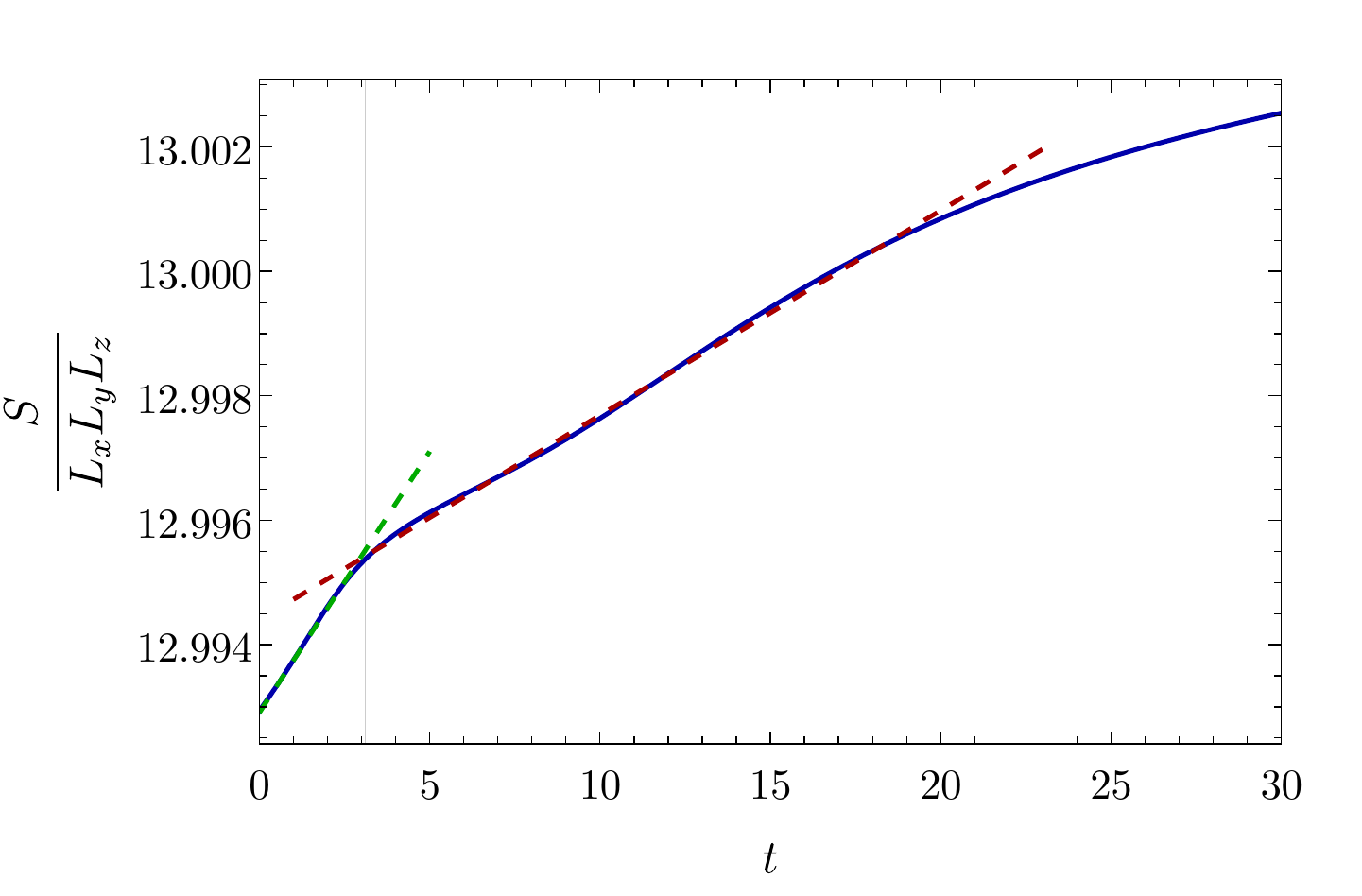}
}
\caption{Average entropy density as a function of time for a \textit{charged oblate} blob collision with $\delta x = 2$. The vertical gray line indicates the collision time $t_c\approx 3.11$. The dashed lines are linear fits, whose slopes are $(8.39\cdot 10^{-4}, 3.29\cdot 10^{-4})$.}
\label{fig:entropy_growth_asymmetric}
\end{figure}

We begin by analyzing the {\it charged oblate} blob collisions. In Figure \ref{fig:entropy_growth_asymmetric} we show the average entropy density (integrated entropy over the three non passive directions and divided by the volume) for the collision with $\delta x = 2$. We can clearly distinguish between three different stages in the collision dynamics. First, we observe a linear growth at early times, before the collision has happened. This growth corresponds to the diffusion of the moving blobs. More details about it will be given below. At $t_c$, marked by a vertical gray line, the slope of the linear growth becomes smaller. In this second stage, lasting for $\Delta t \approx 15$, the system continues producing entropy in a linear fashion. This time, however, on top of the linear growth, there is now a slight oscillation. Notice that both the hydrodynamization and isotropization times fall inside this post-collision stage. At late times, $t \geq 20$, the entropy clearly departs from the linear growth and its rate of production slows down once again. This last stage is likely to be the longest since the equilibrium value for the entropy is $s_\text{final} \approx 13.0082$, which means that half of the total entropy jump is still to be produced, but at a lower rate. An analogous linear entropy growth was observed in planar-shockwave collisions at finite $D$ in \cite{Muller:2020ziz}. The end time of our simulations is large enough to observe the eventual departure from linearity that could not be observed there.
\begin{figure}[thpb]
\centerline{
\includegraphics[width=0.8\textwidth]{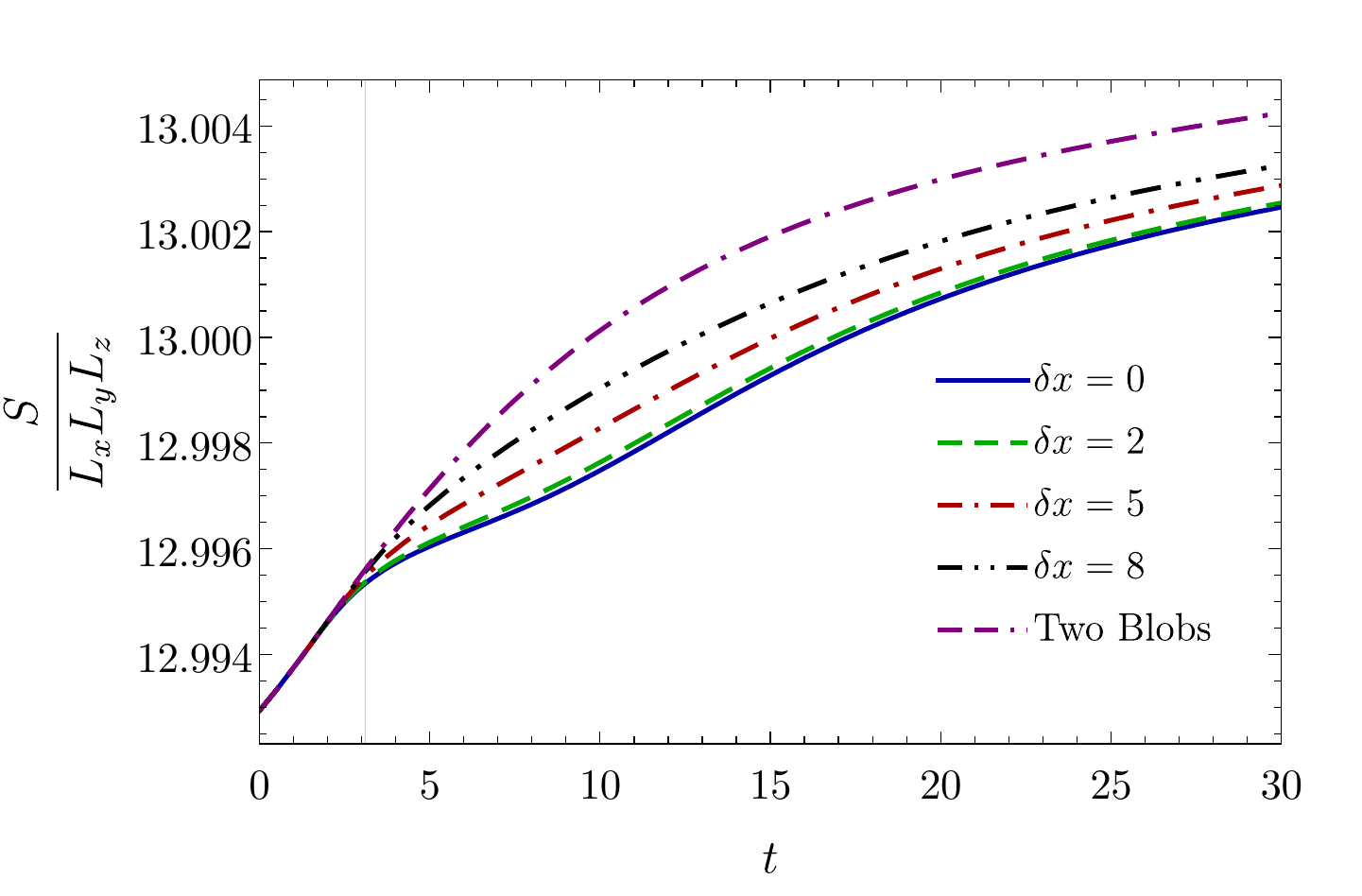}
}
\caption{Average entropy density as a function of time for \textit{charged oblate} blob collision with several different impact parameters $\delta x$. The vertical gray line indicates the collision time, $t_c\approx 3.11$. As a reference, the entropy for the motion of two non-colliding \textit{charged oblate} blobs is included.}
\label{fig:entropy_growth_asymmetric_dx_dependence}
\end{figure}
The general qualitative features observed in Figure \ref{fig:entropy_growth_asymmetric}, including the linear growths and the presence of three stages, can be identified in most collisions with charge. 

A comparison of the entropy time dependence for the {\it charged oblate} collisions is shown in Figure \ref{fig:entropy_growth_asymmetric_dx_dependence}. As a reference, we have included twice the entropy of a single moving {\it charged oblate} blob. At early times, before $t_c \approx 3.11$, all the curves coincide. This fact proves that the pre-collision entropy growth can be understood as coming from the diffusion of two freely moving {\it charged oblate} blobs on the background horizon. For times near $t_c$, differences start to arise. For small values of the impact parameter, $\delta x  \leq 5$, we can still identify a second stage of linear entropy growth. The values we measured for the slopes are $3.22\cdot 10^{-4}$, $3.29\cdot 10^{-4}$ and $3.56\cdot 10^{-4}$ for $\delta x = 0, 2$ and $5$ respectively. The similarity among the values suggests a possible insensitivity to the initial data details and a dependence only on the final, equilibrium state. As $\delta x$ is increased, the length of the second stage decreases until it fully disappears. As expected, higher impact parameters exhibit an entropy behavior that is more similar to that of two freely moving blobs.

The results also show that the collision dynamics slows down the entropy generation with respect to the diffusion of freely moving blobs. Contrary to what we expected, the highest entropy generation rate does not happen for head-on collisions. Higher impact parameters imply higher entropy rates during the second stage. In particular, freely moving blobs generate entropy faster than the complicated dynamics in collisions (for the time window presented here).

\begin{figure}[thpb]
\centerline{
\includegraphics[width=0.8\textwidth]{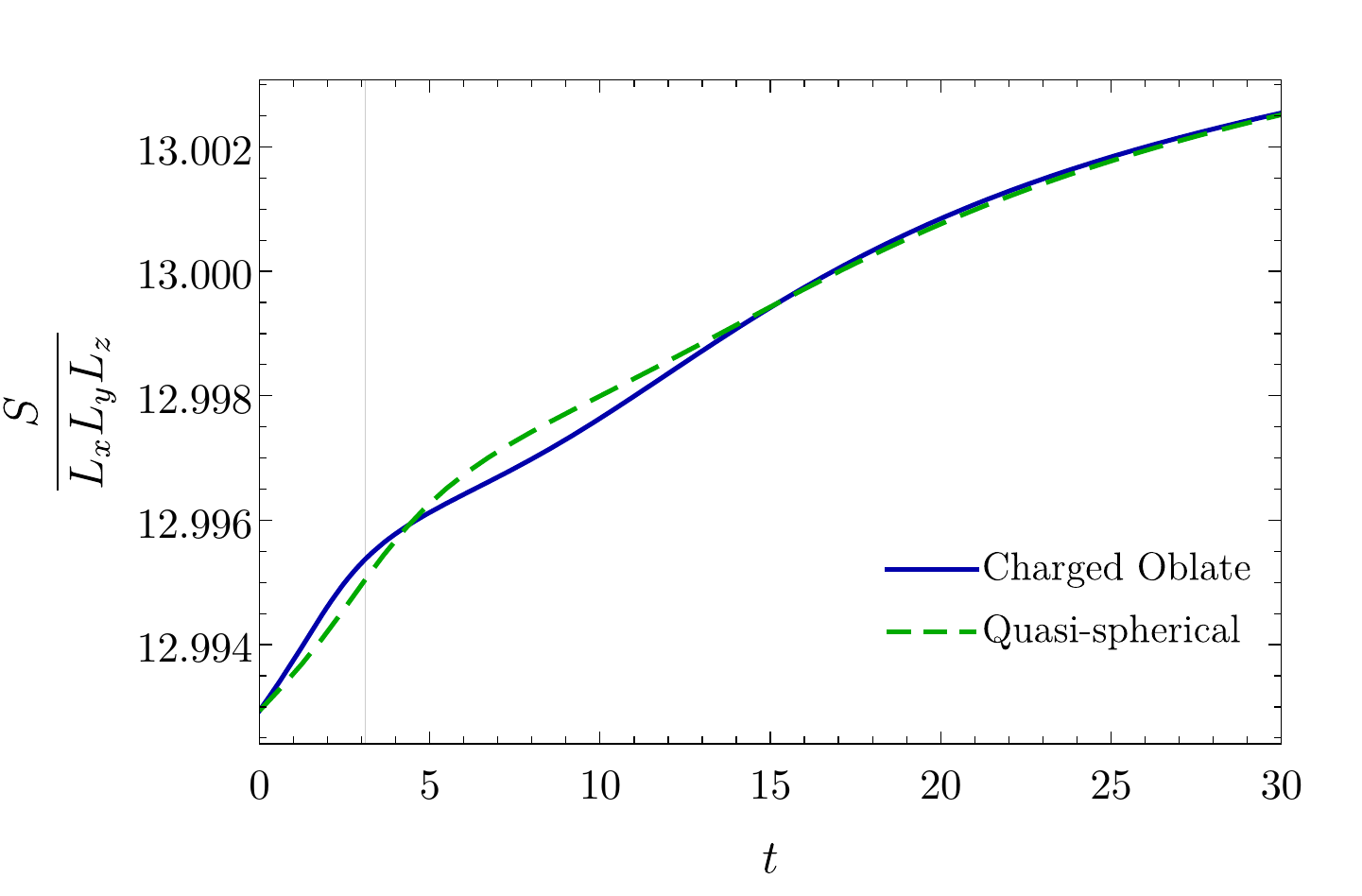}
}
\caption{Total entropy normalized by the volume as a function of time for a \textit{charged oblate} with $\delta x = 2$ and \textit{quasi-spherical} blob collisions. The width of the initial blobs is chosen in such a way that the end state is the same. The measured slopes are $(3.29\cdot 10^{-4}, 2.86\cdot 10^{-4})$}
\label{fig:entropy_growth_asymmetric_vs_symmetric}
\end{figure}

In order to decide whether the measured rates in Figure \ref{fig:entropy_growth_asymmetric_dx_dependence} are insensitive to all the details of the initial data, in Figure \ref{fig:entropy_growth_asymmetric_vs_symmetric} we compare the entropy of a {\it charged oblate} blob collision with a {\it quasi-spherical} blob collision, both with $\delta x = 2$. Both setups also have identical total mass and charge values, which means that the final equilibrium state is the same. The only difference is on the initial blob shapes. The measured rates are $3.29\cdot 10^{-4}$ for the {\it charged oblate} blob collision, and $2.86\cdot 10^{-4}$ for the {\it quasi-spherical} blob one. We observe a bigger discrepancy between their values than for collisions of identical blobs with different impact parameters. 

The linear growth rate in the post-collision stage is not independent of the details of the initial data. Interestingly, at later times both curves behave in a similar way, which seems to indicate that the details of the initial data have been forgotten.
We also studied the time evolution of the entropy for blobs with different values of the charge $\delta q$, and observed that the qualitative behavior of the entropy is maintained, and all three stages can be identified for small enough impact parameters.

\subsection{Neutral collisions}

\begin{figure}[thpb]
\centerline{
\includegraphics[width=0.8\textwidth]{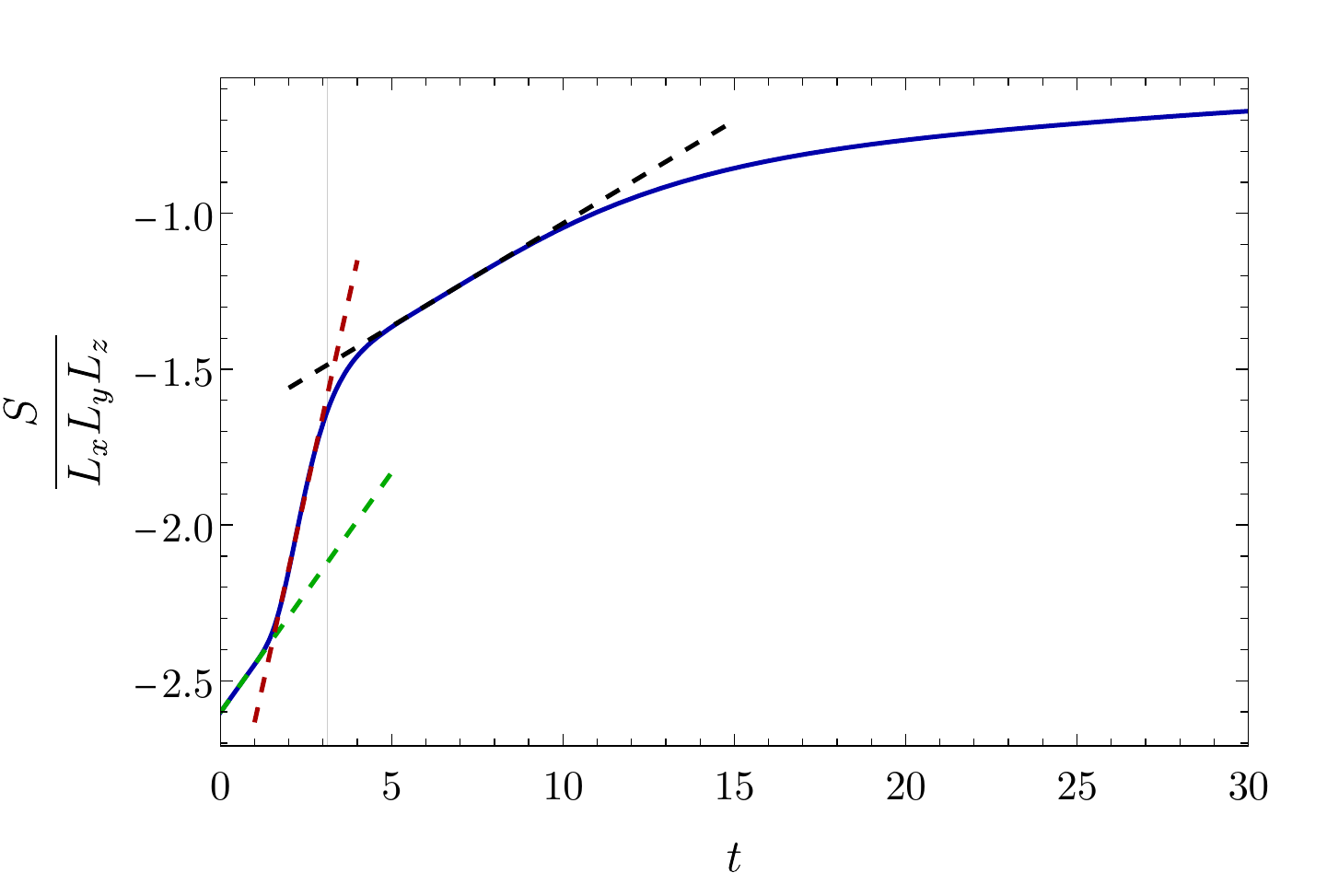}
}
\caption{Total entropy normalized by the volume as a function of time for an \textit{neutral oblate} blob collision with $\delta x = 2$. The dashed lines correspond to linear fits, whose slopes are $(0.1541, 0.4945, 0.0660)$. The first change in the slope takes place at $t \approx 1.5$, while the vertical gray marks the collision time $t_c\approx 3.11$.}
\label{fig:entropy_growth_asymmetric_neutral}
\end{figure}

\begin{figure}[thpb]
\centerline{
\includegraphics[width=0.5\textwidth]{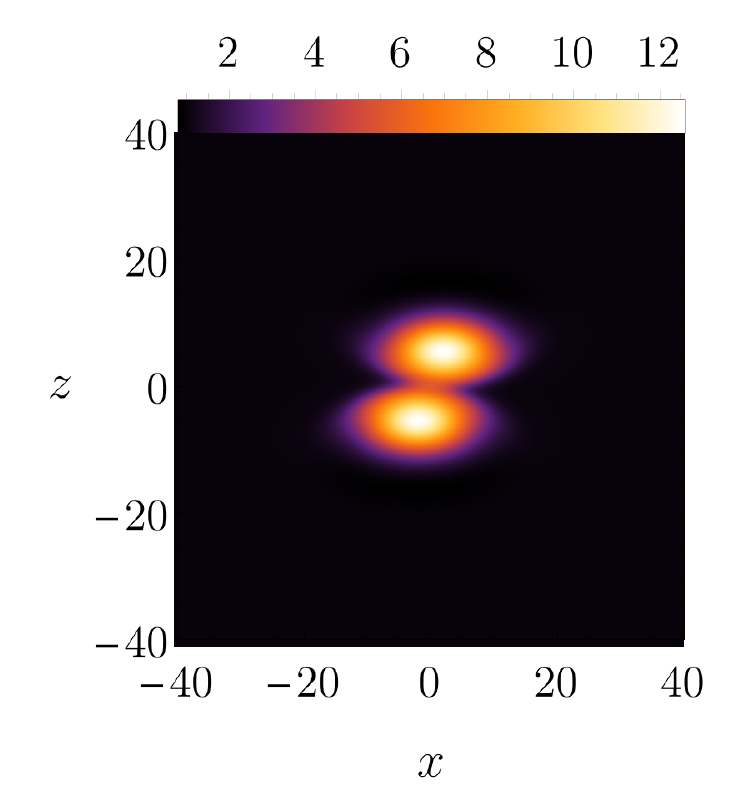}
}
\caption{Mass density at $t = 1.5$ at the collision plane. At this instant the two blobs start to considerably overlap each other and the collision process starts.}
\label{fig:1st_contact}
\end{figure}

In Figure \ref{fig:entropy_growth_asymmetric_neutral} we show the full evolution of the entropy in a {\it neutral oblate} blob collision with $\delta x = 2$.
In this case, we can identify four different stages of entropy growth, one more than in charged collisions. At early times, we find a linear growth of the entropy which is related to the dissipation of the blobs before they collide. At around $t\approx 1.5$, before $t_c\approx 3.11$ (vertical dashed line), the system enters a second stage of faster linear growth. This phase is completely absent in the charged case. By looking into the mass density at $t \approx 1.5$, we can relate this early change in the rate to the instant in which the blobs start to considerably overlap with each other, as shown in Figure \ref{fig:1st_contact}. Indeed, at $t\approx1.5$, the mass density at the origin is about half the maximum of the blobs. This stage lasts from $t \approx 1.5$ to $t_c$, and its entropy growth rate is the fastest of the whole evolution. The fact that this second stage is missing in the evolution for charged blobs indicates that the entropy \eqref{eq:entropy_neutral} is more sensitive to the presence of new dynamical regimes.
After $t_c$, the entropy enters even a new regime of linear growth, this time with a smaller slope. This stage is analogous to the second regime of linear growth found in Figure \ref{fig:entropy_growth_asymmetric}, but with about a third of the duration ($\Delta t\approx 5$). At $t \approx 10$, the entropy production stops following a linear trend and its growth rate decreases. Contrary to the charged case, most of the expected entropy increment has already taken place. In fact, by the end of our simulation at $t=30$, the total increase in the entropy has been of around $s_ \text{final}-s_\text{initial}\approx 1.9$, about 90\% of the expected total jump.  
\begin{figure}[thpb]
\centerline{
\includegraphics[width=0.8\textwidth]{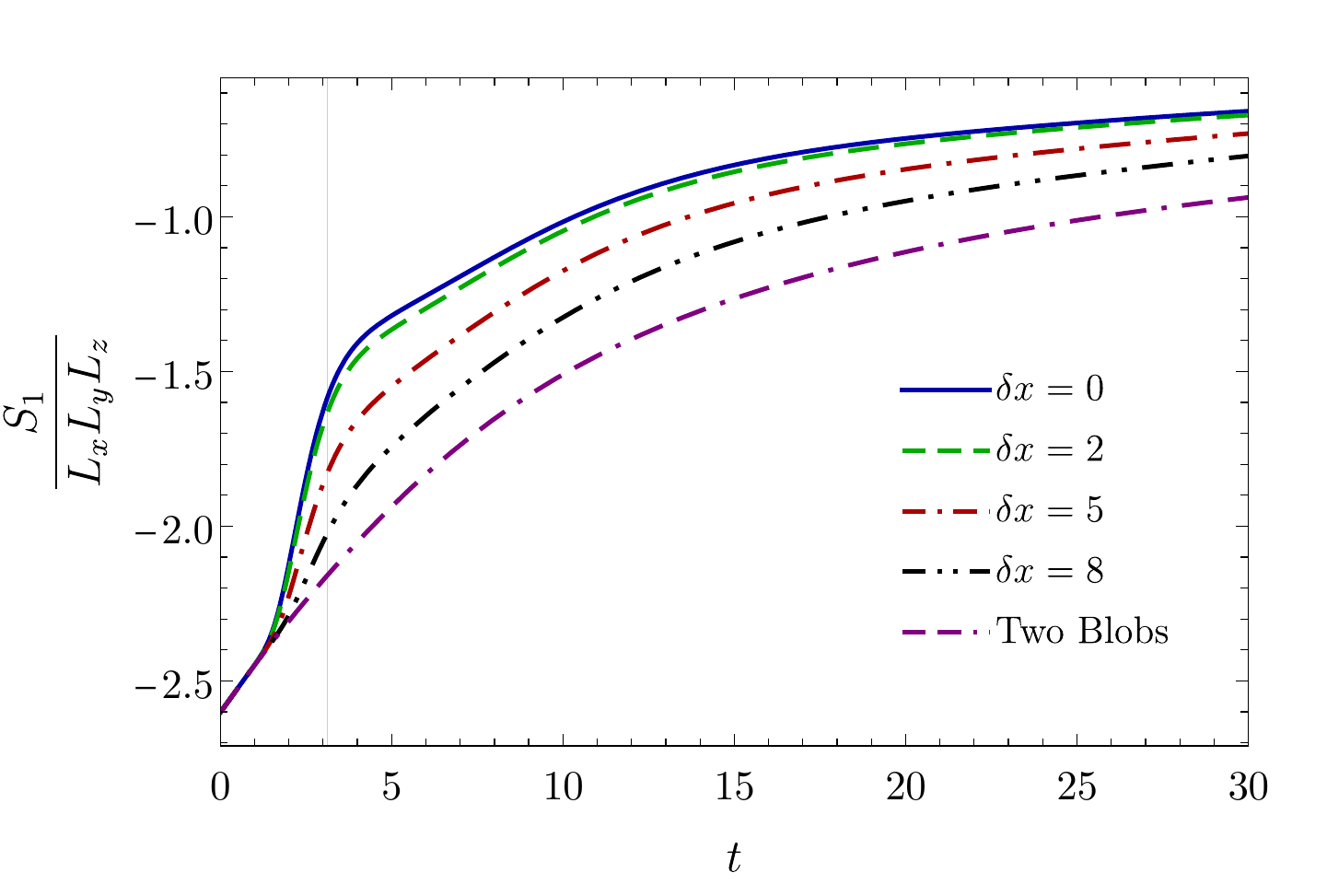}
}
\caption{Total entropy normalized by the volume as a function of time for identical initial data with different impact parameters. The vertical gray line indicates the collision time. As a reference, the entropy for the motion of two non-colliding blobs is included.}
\label{fig:entropy_growth_asymmetric_dx_dependence_neutral}
\end{figure}

Finally, in Figure \ref{fig:entropy_growth_asymmetric_dx_dependence_neutral} we show the entropy of {\it neutral oblate} blob collisions. We also add the entropy produced by two freely diffusing blobs. As the impact parameter is increased, the curves resemble more the non-colliding blobs, just as expected. More intuitively than in the charged setup, the collision accelerates the entropy growth thanks to the second stage, which is absent in charged collisions.

We conclude that the first stage of linear growth is due to the diffusion of the blobs before they get into contact, while the second stage clearly depends on the impact parameter. This is seen in the slope values of $0.5233$, $0.4945$ and $0.3731$ when taking impact parameters of $\delta x = 0$, $2$ and $5$ respectively. Regarding the slope of the linear entropy production after $t_c$, we measured clear dependence on the impact parameter, with slopes of $0.0627$, $0.0660$ and $0.0777$ for $\delta x = 0$, $2$ and $5$. The results therefore show that the second and third linear growth rates do depend on the impact parameter.

\section{Discussion}
\label{sec:Discussion}

Although the AdS/CFT correspondence transforms problems that are hard to address within Quantum Field Theory into tractable classical gravity ones, the computational cost of solving Einstein's equations without further assumptions is still large. In the present work we used the large $D$ limit of General Relativity to drastically simplify the problem of shockwave collisions, allowing us to scan over different kinds of collision scenarios with full 4+1-dimensional dynamics. We provided an overview of collisions of different blob sizes, shapes, charges and impact parameters. However, the formalism has its own limitations. The effective description \eqref{eq:Large-D-effecttive-equations} becomes non-relativistic, including the equation of state, and the transport coefficients differ from those in 4+1 dimensions. Presumably, these have little effect at the qualitative level. The most important limitation is that the background horizon temperature cannot be parametrically suppressed while keeping the blob amplitude fixed, which produces undesired dissipative effects.

In a similar way to what is observed in AdS$_5$ collisions, the system produces a large amount of anisotropy as well as deviations from first-order hydrodynamics. In other words, the second order gradient in \eqref{eq:constitutive-relations} plays an important role around the collision time. After a few units of time, the system relaxes back to a nearly isotropic state which is well captured by first order hydrodynamic terms. Contrary to what has been observed in AdS$_5$ collisions \cite{Chesler:2010bi, Casalderrey-Solana:2013aba, Chesler:2015wra}, the hydrodynamization and isotropization times approximately coincide at large $D$.

As the impact parameter is increased, the maximum mass, charge density and anisotropty decrease. We found that, for impact parameters below the blobs' width, the evolution of the mass and charge densities are the same when normalized by their maximum value. By colliding blobs of increasing charge values, we conclude that the charge only plays an important role for the entropy, but not for the rest of variables. A similar weak effect on the rest of observables was observed in finite $D$ planar shockwave collisions \cite{Casalderrey-Solana:2016xfq}. It is unknown to us if larger values of blob charge would change our conclusions.

We additionally studied the entropy produced during collisions. At large $D$, entropy generation by viscous dissipation is $1/D$ suppressed over charge diffusion. Hence, one has to consider two different notions of entropy for charged and neutral collisions. For both kinds of collisions, we observed several regimes of entropy production linear in time. 

When the incoming blobs are charged, the entropy grows linearly in time at different rates before and after the collision. Eventually, a departure from linear growth is observed. Interestingly, the growth rate after the collision increases with the impact parameter. In neutral collisions, we find an extra linear growth stage. This stage starts when the blobs start overlapping with each other and ends at the collision time. The highest rate of entropy production occurs at this stage, although such rate decreases with increasing impact parameter.

Similar post-collision linear entropy growths were observed at finite $D$ collisions \cite{Muller:2020ziz, Grozdanov:2016zjj}. In \cite{Muller:2020ziz} a connection between the growing rate and the largest Lyapunov exponent was suggested through the Kolmogorov-Sinai (KS) entropy. Even if certain notions of entropy may exhibit regimes of linear growth whose rate is equal to the KS entropy, e.g. \cite{KS_vs_physical, Kunihiro:2008gv}, the entropy definitions used here may not. In fact, KS entropy is only sensitive to the end state, while the rates we observed depend on the initial data. Given that our boundary theory has a classical gravitational dual, the maximum Lyapunov exponent saturates the Maldacena-Shenker-Stanford bound $\lambda_L\leq2\pi \boldsymbol{T}$ \cite{Maldacena:2015waa}, with $\boldsymbol{T}$ the physical temperature, which diverges as $D\rightarrow\infty$. One can presumably recover the bound by working with appropriately rescaled quantities. We wonder whether the large $D$ limit can simplify the study of chaos too. Progress in this direction will be reported in \cite{DMM}. 

Linear entropy growths were also observed in the context of holographic collisions of phase domains in a theory with a first-order thermal phase transition \cite{Bea:2021ieq}. This corresponds to a completely different type of setup, which suggests that stages of linear entropy production are a signature in collision dynamics.

It would be interesting to investigate the effects of $1/D$ corrections, which would allow us to more precisely observe the deviations that finite $D$ introduce into the results presented here, even if only perturbatively. The resulting equations would increase in difficulty, however they would nevertheless still represent a major simplification to AdS$_5$ collisions.

\section*{Acknowledgements}
\label{sec:Acknowledgements}

We thank David Ramirez and Martin Sasieta for useful discussions. We are grateful to David Mateos and Roberto Emparan and Marija Toma{\v s}evi\'c for their very useful comments on the first manuscript. We thank David Licht, Ryotaku Suzuki and Roberto Emparan for their early exploratory work in this line of research.
RL acknowledges financial support provided by Next Generation EU through a University of Barcelona Margarita Salas grant from the Spanish Ministry of Universities under the {\it Plan de Recuperaci\'on, Transformaci\'on y Resiliencia} and by Generalitat Valenciana / Fons Social Europeu through APOSTD 2022 post-doctoral grant CIAPOS/2021/150. Work supported by Spanish Agencia Estatal de Investigaci\'on (Grant PID2021-125485NB-C21) funded by MCIN/AEI/10.13039/501100011033 and ERDF A way of making Europe, and the Generalitat Valenciana (Grant PROMETEO/2019/071).
The work of MSG is supported by the European Research Council (ERC) under the European Union's Horizon 2020 research and innovation program (grant agreement No758759).

\appendix

\section{Quasinormal modes and charged silence}
\label{app:QNM}

As a first test of the numerical code, we perform an analysis of the large $D$ quasinormal spectrum of the AdS black branes, focusing on the sound channel. Introducing a Fourier perturbation to the uniform brane solution in the form 
\begin{equation}
\begin{split}
    \rho &= \rho_0 + \delta \rho \,e^{-i\omega t + i k_j x^j},\\
    q &= q_0 + \delta q \,e^{-i\omega t + i k_j x^j},\\
    p^i &= \delta p^i\, e^{-i\omega t + i k_j x^j},
\end{split}
\end{equation}
one obtains an eigenvalue problem for $\delta \rho$, $\delta q$ and $\delta p^i$, whose eigenvalues and eigenvectors we can classify into charge diffusion, shear and sound modes \cite{Emparan:2016sjk}. In order to interpret the nature of each family of modes, it is illustrative to look at the linear span generated by the eigenvectors $(\delta \rho, \delta q;  \delta p^i)$ of each eigenvalue $\omega$. Let us consider a system with 3+1 dependence, and assume without loss of generality that the 3-vector $\vec k$ is aligned with the $z$-axis, i.e., $\vec k = (0, 0, k)$. Then, we obtain:

\begin{itemize}
    \item  Charge diffusion mode: 
    \begin{equation}
    \begin{split}
        \omega =& -i k^2, \\
        (\delta \rho, \delta q; \delta \vec p) \, : \; &(0,\; 1;\;0,\;0,\;0). 
    \end{split}
    \end{equation}
    \item  Shear modes: 
    \begin{equation}
    \begin{split}
        \omega =& -i a_+ k^2, \\
        (\delta \rho, \delta q; \delta \vec p) \, : \; &(0,\; 0;\; 1,\; 0,\; 0), \\
                                                           &(0,\; 0;\; 0,\; 1,\; 0). 
    \end{split}
    \end{equation}
    \item  Sound modes: 
    \begin{equation}
    \begin{split}
        \omega =& \pm k \sqrt{1 - k^2 a_-^2} - i a_+ k^2, \quad (\delta \rho, \delta q; \delta \vec p) \, : \\ \; \left(\sqrt{1 - k^2 a_-^2},\; \right.&\left.  \frac{q_0}{\rho_0}\sqrt{1 - k^2 a_-^2};\; \pm 1,\; \pm 1,\;  i a_- k \sqrt{1 - k^2 a_-^2} \pm (k^2 a_-^2 - 1) \right),
    \end{split}
    \label{eq:sound_qnm}
    \end{equation}
\end{itemize}
where we have defined
\begin{equation}
    a_\pm = \frac12 \left( 1 \pm \sqrt{1-\frac{2q_0^2}{\rho_0^2}}\right).
\end{equation}
A surprising feature of the sound mode frequency is that it becomes purely imaginary for $k > \frac{1}{a_-}$, a phenomenon which was given the name of {\it charged silence} in \cite{Emparan:2016sjk}. Perturbations with high wavenumber do not propagate, only diffuse. In the particular case when $k = \frac{1}{a_-}$, we can see from \eqref{eq:sound_qnm} that the sound mode reduces to a linear combination of shear modes. Additionally, it is important to notice that charge propagation via the sound channel will only occur when the background charge $q_0$ is different from zero, as shown in \eqref{eq:sound_qnm}. Otherwise, the sound mode perturbation will necessarily have $\delta q = 0$ and only the charge diffusion mode will be excited in the charge density field. Even though this effect is seen here at the linear level, it seems that a similar feature is observed at the nonlinear level, as seen in section \ref{sec:Collisions}.
\begin{figure}[thpb]
\centerline{\includegraphics[width=0.6\textwidth]{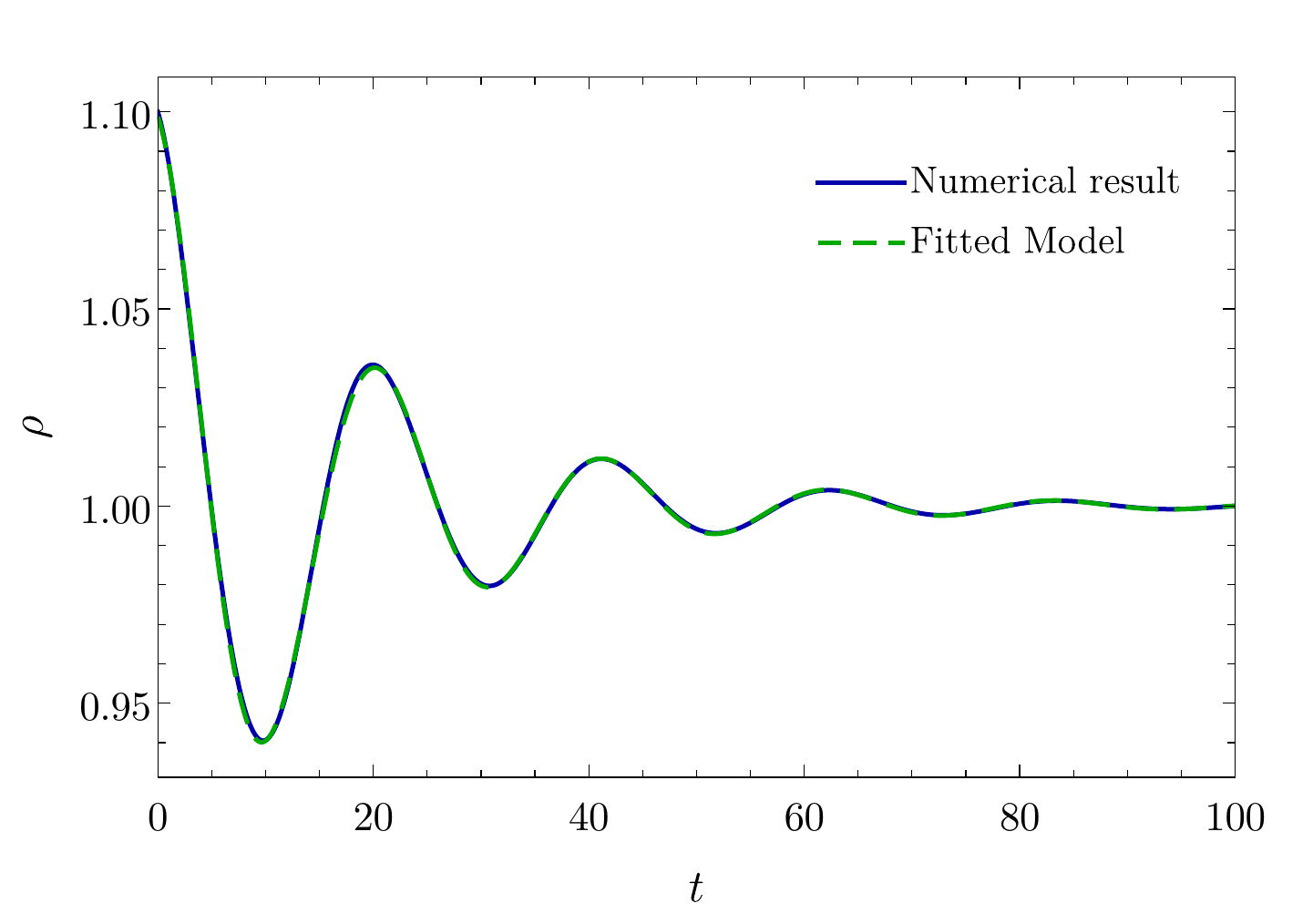}}
\caption{Time dependence of the mass density $\rho$ at the origin, for a sound mode perturbation with $k = 0.3$ and $q_0/\rho_0 = 0.7$, together with the fitted model of the form \eqref{eq:sinusoid_model}. 
\label{fig:damped_sinusoid}}
\end{figure}
\begin{figure}[thpb]
\centerline{\includegraphics[width=0.7\textwidth]{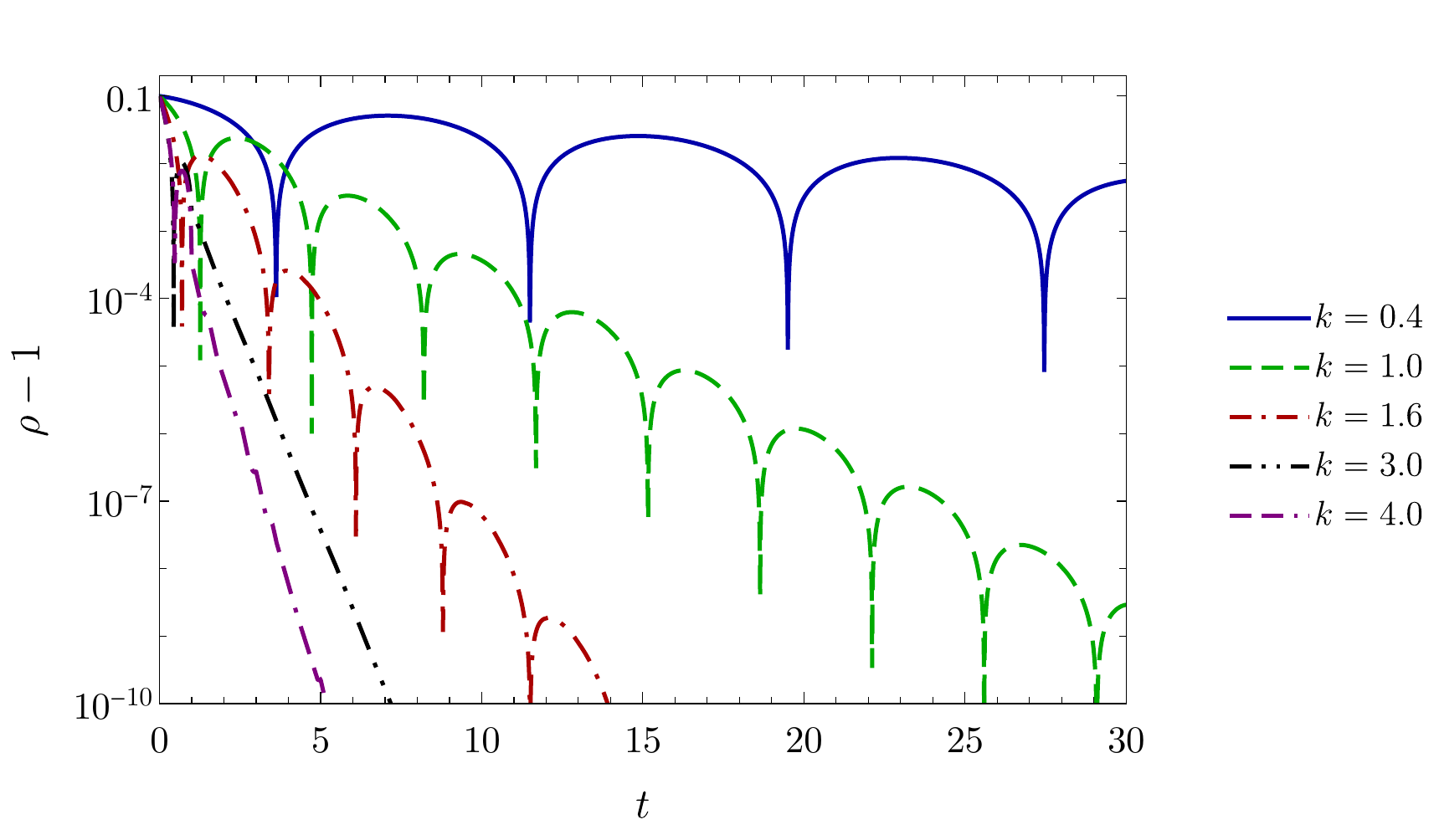}}
\caption{Quasinormal sound ringdown profiles for $q_0/\rho_0 = 0.7$ at several values of the wavenumber $k$. The phenomenon of charged silence can be clearly seen, as the modes with $k > 1/a_- \approx 2.33$ do not propagate and simply decay exponentially.
\label{fig:log_modes}}
\end{figure}
\begin{table}[thpb]
    \centering
    \begin{tabular}{|c|c|c|}
    \hline
        $k$ & Analytical $\omega$ & Fitted $\omega$ \\ \hline
        0.4 & 0.3941 - 0.0913 $i$ &  0.3962 - 0.0907 $i$ \\ \hline
        1.0 & 0.9032 - 0.5707 $i$ &  0.9041 - 0.5723 $i$ \\ \hline
        1.5 & 1.1629 - 1.4610 $i$ &  1.1664 - 1.4577 $i$ \\ \hline
        3.0 & - 2.7018 $i$ &  - 2.6991 $i$ \\ \hline
        4.0 & - 3.5476 $i$ &  - 3.5334 $i$ \\ \hline
    \end{tabular}
    \caption{Comparison between the analytical sound frequencies from \eqref{eq:sound_qnm} and the fitted values of Figure (\ref{fig:log_modes}). \label{tab:mode_fits}}
\end{table}
As a first test for the numerical code, we will focus here on the sound channel, which contains a richer phenomenology than the other large $D$ channels (shear and charge diffusion). In order to test the sound mode in the numerical simulations, we will use $z$-dependent initial data designed to excite the sound mode, of the form
\begin{equation}
\begin{split}
    \rho(t = 0) &= \rho_0 + \epsilon \cos(z k), \\
    q(t = 0) &= q_0 + \left(\frac{q_0}{\rho_0}\right) \epsilon \cos(z k), \\
    p^i(t=0) &= 0,
\end{split}
\end{equation}
and we choose a setup close to extremality $(\rho_0, q_0) = (1, 0.7)$, and $\epsilon = 0.1$. For this background charge, we expect charged silence to appear at $k \gtrsim 2.33$. In the particular case of the quasinormal modes, we will only have dependence on the $z$ direction, so we will simply choose $N_x = N_y = 2$ and $N_z = 64$. Also, testing different values of $k$ requires us to vary the size of the computational domain, while keeping a sufficiently small time step in order to make sure that the Courant–Friedrichs–Lewy (CFL) condition is always satisfied. For this reason, we will take
\begin{equation}
    L_z = \frac{20\pi}{k}, \; \Delta t = 0.01.
\end{equation}
The values of $L_x$ and $L_y$ are irrelevant in this case. By plotting the value of $\rho$ at the origin as a function of time, we can clearly extract the value of $\omega = \omega_R + i \omega_I$ from the oscillation frequency and damping rate by numerically fitting a function of the form
\begin{equation}
    \rho(t) = A e^{\omega_I t} \cos(\omega_R t + \phi).
    \label{eq:sinusoid_model}
\end{equation}
\begin{figure}[thpb]
\centerline{\includegraphics[width=0.6\textwidth]{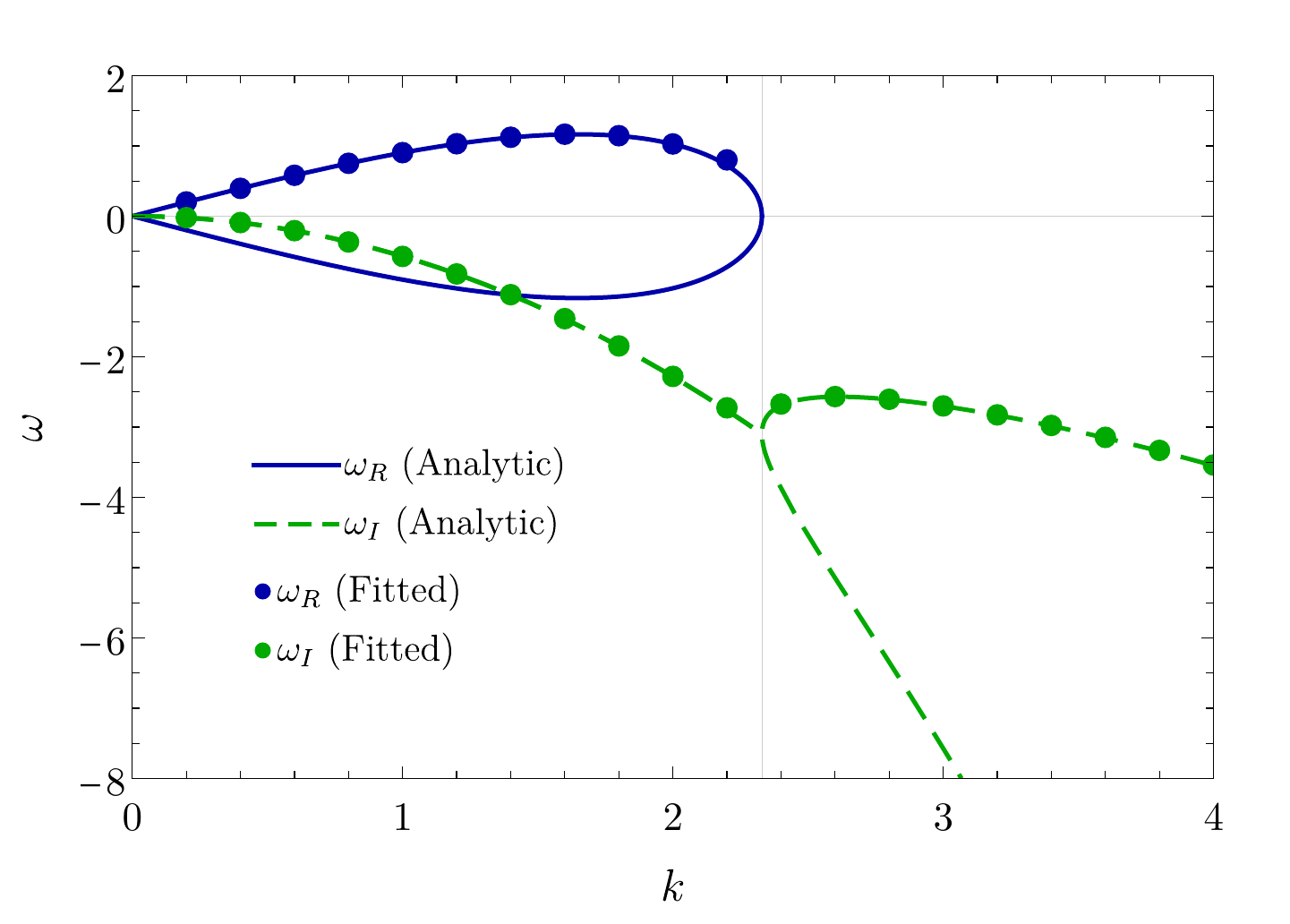}}
\caption{Values of the real (blue) and imaginary (green) parts of the quasinormal sound frequency at $q_0/\rho_0 = 0.7$. The continuous lines correspond to the analytic prediction of \eqref{eq:sound_qnm}, while the points mark the numerical values fitted from the code output. The charged silence threshold is highlighted by the vertical line.
\label{fig:charged_silence}}
\end{figure}
An example of the procedure is depicted in Figure \ref{fig:damped_sinusoid}, for $k = 0.3$, showing the characteristic damped sinusoid profile of quasinormal ringdown. Figure \ref{fig:log_modes} shows the perturbation decaying in a logarithmic scale, for several values of $k$. As expected the modes satisfying $k \gtrsim 2.33$ exhibit charged silence by exponentially decaying without oscillation. The analytical values of the frequencies, extracted from equation \eqref{eq:sound_qnm}, are compared to the numerically fitted values in Table \ref{tab:mode_fits}. A more exhaustive fitting of the quasinormal sound modes is displayed in Figure \ref{fig:charged_silence} for values of $k$ between 0 and 4. For values of $k$ above the silence threshold, there exist two different imaginary modes. In the numerical setup we will observe only the dominant one, with the smallest absolute value.

\section{Convergence testing}
\label{app:convergence}

In order to assess the quality of the numerical code, we have performed a self-convergence test, by comparing solutions with different resolutions. 
\begin{figure}[h!]
\centerline{\includegraphics[width=0.6\textwidth]{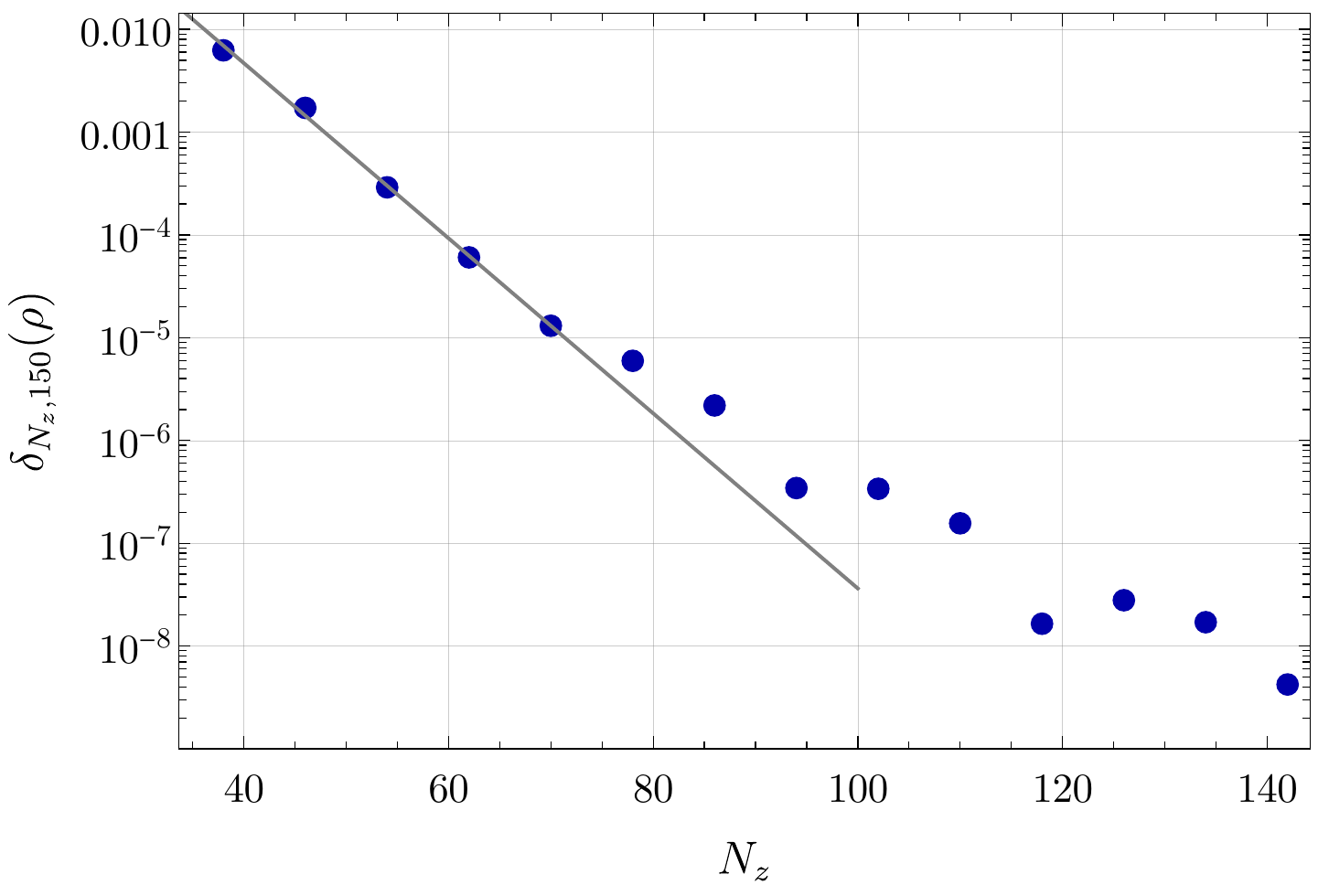}}
\caption{$\delta_{N_z,150}(\rho)$ for the $z=0$ plane at the collision time $t=3$ as a function o the number of points in the $z$ direction. The gray line is a fit to the first five points, with equation $11.96 \; e^{-0.196\; N_z}$.
\label{fig:convergence_testing}}
\end{figure}
We have done so in the particular case of an off-centered collision similar to the one in Figure \ref{fig:snapshots_asymmetric} but with a box of sizes $L_x = L_y = L_z = 50$. By keeping fixed the time step and the number of points in the normal directions $N_x = N_y = 50$, we vary the value of $N_z$ from 38 to 150 at intervals of 8. We define the quantity $\delta_{N,M}(f)$ as the $l_\infty$ norm of the relative error between the values of the grid function $f$ under different number of computational points $N$ and $M$. 
\begin{equation}
    \delta_{N,M}(f) = \max \left| 1 - \frac{f_N}{f_M}\right|
\end{equation}
In our case, we compare the values of $\rho$ the points in the collision plane, i.e., $z=0$ at the approximate time of collision $t = 3$. Due to the symmetry of the computational grid, the points at the collision plane remain fixed when we change $N_z$, thus allowing us to compare them without the need of any interpolation. In Figure \ref{fig:convergence_testing} we plot $\delta_{N_z,150}(\rho)$ in a logarithmic scale as a function of $N_z$. In this case we are taking the run with $N_z = 150$ as the most accurate reference, and comparing the other simulations to this case. From the plot we can see that the points are approximately lying on a straight line, showing that the code convergence is exponential, as expected for a spectral grid. An exponential fit to the first points reveals a trend of the form $\delta_{N_z,150}(\rho) = 11.96 \; e^{-0.196\; N_z}$. For $N_z \gtrsim 70$ the convergence rate decreases, possibly by truncation error, but does not deviate dramatically from the fitted line. The fact that there is still convergence at such a high number of spectral points is concordant with the observation that the simulations still improve at least up to $N_z = 150$. 

\bibliography{refs}{}

\providecommand{\href}[2]{#2}\begingroup\raggedright\begin{thebibliography}{10}

\bibitem{Nastase:2005rp}
H.~Nastase, {\it {The RHIC fireball as a dual black hole}},
  \href{http://arxiv.org/abs/hep-th/0501068}{{\tt hep-th/0501068}}.

\bibitem{Janik:2005zt}
R.~A. Janik and R.~B. Peschanski, {\it {Asymptotic perfect fluid dynamics as a
  consequence of Ads/CFT}},  {\em Phys. Rev. D} {\bf 73} (2006) 045013,
  [\href{http://arxiv.org/abs/hep-th/0512162}{{\tt hep-th/0512162}}].

\bibitem{Janik:2006gp}
R.~A. Janik and R.~B. Peschanski, {\it {Gauge/gravity duality and
  thermalization of a boost-invariant perfect fluid}},  {\em Phys. Rev. D} {\bf
  74} (2006) 046007, [\href{http://arxiv.org/abs/hep-th/0606149}{{\tt
  hep-th/0606149}}].

\bibitem{Kovchegov:2007pq}
Y.~V. Kovchegov and A.~Taliotis, {\it {Early Time Dynamics in Heavy Ion
  Collisions from AdS/CFT Correspondence}},  {\em Phys. Rev. C} {\bf 76} (2007)
  014905, [\href{http://arxiv.org/abs/0705.1234}{{\tt arXiv:0705.1234}}].

\bibitem{Grumiller:2008va}
D.~Grumiller and P.~Romatschke, {\it {On the collision of two shock waves in
  AdS(5)}},  {\em JHEP} {\bf 08} (2008) 027,
  [\href{http://arxiv.org/abs/0803.3226}{{\tt arXiv:0803.3226}}].

\bibitem{Lin:2009pn}
S.~Lin and E.~Shuryak, {\it {Grazing Collisions of Gravitational Shock Waves
  and Entropy Production in Heavy Ion Collision}},  {\em Phys. Rev. D} {\bf 79}
  (2009) 124015, [\href{http://arxiv.org/abs/0902.1508}{{\tt
  arXiv:0902.1508}}].

\bibitem{Beuf:2009cx}
G.~Beuf, M.~P. Heller, R.~A. Janik, and R.~Peschanski, {\it {Boost-invariant
  early time dynamics from AdS/CFT}},  {\em JHEP} {\bf 10} (2009) 043,
  [\href{http://arxiv.org/abs/0906.4423}{{\tt arXiv:0906.4423}}].

\bibitem{Kovchegov:2009du}
Y.~V. Kovchegov and S.~Lin, {\it {Toward Thermalization in Heavy Ion Collisions
  at Strong Coupling}},  {\em JHEP} {\bf 03} (2010) 057,
  [\href{http://arxiv.org/abs/0911.4707}{{\tt arXiv:0911.4707}}].

\bibitem{Gubser:2009sx}
S.~S. Gubser, S.~S. Pufu, and A.~Yarom, {\it {Off-center collisions in AdS(5)
  with applications to multiplicity estimates in heavy-ion collisions}},  {\em
  JHEP} {\bf 11} (2009) 050, [\href{http://arxiv.org/abs/0902.4062}{{\tt
  arXiv:0902.4062}}].

\bibitem{Romatschke:2013re}
P.~Romatschke and J.~D. Hogg, {\it {Pre-Equilibrium Radial Flow from Central
  Shock-Wave Collisions in AdS5}},  {\em JHEP} {\bf 04} (2013) 048,
  [\href{http://arxiv.org/abs/1301.2635}{{\tt arXiv:1301.2635}}].

\bibitem{Bantilan:2018vjv}
H.~Bantilan, T.~Ishii, and P.~Romatschke, {\it {Holographic Heavy-Ion
  Collisions: Analytic Solutions with Longitudinal Flow, Elliptic Flow and
  Vorticity}},  {\em Phys. Lett. B} {\bf 785} (2018) 201--206,
  [\href{http://arxiv.org/abs/1803.10774}{{\tt arXiv:1803.10774}}].

\bibitem{Kajantie:2008rx}
K.~Kajantie, J.~Louko, and T.~Tahkokallio, {\it {Gravity dual of conformal
  matter collisions in 1+1 dimensions}},  {\em Phys. Rev. D} {\bf 77} (2008)
  066001, [\href{http://arxiv.org/abs/0801.0198}{{\tt arXiv:0801.0198}}].

\bibitem{Albacete:2009ji}
J.~L. Albacete, Y.~V. Kovchegov, and A.~Taliotis, {\it {Asymmetric Collision of
  Two Shock Waves in AdS(5)}},  {\em JHEP} {\bf 05} (2009) 060,
  [\href{http://arxiv.org/abs/0902.3046}{{\tt arXiv:0902.3046}}].

\bibitem{Chesler:2010bi}
P.~M. Chesler and L.~G. Yaffe, {\it {Holography and colliding gravitational
  shock waves in asymptotically AdS$_{5}$ spacetime}},  {\em Phys. Rev. Lett.}
  {\bf 106} (2011) 021601, [\href{http://arxiv.org/abs/1011.3562}{{\tt
  arXiv:1011.3562}}].

\bibitem{Casalderrey-Solana:2013aba}
J.~Casalderrey-Solana, M.~P. Heller, D.~Mateos, and W.~van~der Schee, {\it
  {From full stopping to transparency in a holographic model of heavy ion
  collisions}},  {\em Phys. Rev. Lett.} {\bf 111} (2013) 181601,
  [\href{http://arxiv.org/abs/1305.4919}{{\tt arXiv:1305.4919}}].

\bibitem{Chesler:2015wra}
P.~M. Chesler and L.~G. Yaffe, {\it {Holography and off-center collisions of
  localized shock waves}},  {\em JHEP} {\bf 10} (2015) 070,
  [\href{http://arxiv.org/abs/1501.04644}{{\tt arXiv:1501.04644}}].

\bibitem{vanderSchee:2015rta}
W.~van~der Schee and B.~Schenke, {\it {Rapidity dependence in holographic heavy
  ion collisions}},  {\em Phys. Rev. C} {\bf 92} (2015), no.~6 064907,
  [\href{http://arxiv.org/abs/1507.08195}{{\tt arXiv:1507.08195}}].

\bibitem{Grozdanov:2016zjj}
S.~Grozdanov and W.~van~der Schee, {\it {Coupling Constant Corrections in a
  Holographic Model of Heavy Ion Collisions}},  {\em Phys. Rev. Lett.} {\bf
  119} (2017), no.~1 011601, [\href{http://arxiv.org/abs/1610.08976}{{\tt
  arXiv:1610.08976}}].

\bibitem{Waeber:2019nqd}
S.~Waeber, A.~Rabenstein, A.~Sch\"afer, and L.~G. Yaffe, {\it {Asymmetric
  shockwave collisions in AdS$_{5}$}},  {\em JHEP} {\bf 08} (2019) 005,
  [\href{http://arxiv.org/abs/1906.05086}{{\tt arXiv:1906.05086}}].

\bibitem{Muller:2020ziz}
B.~M\"uller, A.~Rabenstein, A.~Sch\"afer, S.~Waeber, and L.~G. Yaffe, {\it
  {Phenomenological implications of asymmetric $AdS_5$ shock wave collision
  studies for heavy ion physics}},  {\em Phys. Rev. D} {\bf 101} (2020), no.~7
  076008, [\href{http://arxiv.org/abs/2001.07161}{{\tt arXiv:2001.07161}}].

\bibitem{Waeber:2022tts}
S.~Waeber and L.~G. Yaffe, {\it {Collision of localized shocks in AdS$_5$ as a
  series expansion in transverse gradients}},
  \href{http://arxiv.org/abs/2206.01819}{{\tt arXiv:2206.01819}}.

\bibitem{Waeber:2022vgf}
S.~Waeber and L.~G. Yaffe, {\it {Colliding localized, lumpy holographic shocks
  with a granular nuclear structure}},
  \href{http://arxiv.org/abs/2211.09190}{{\tt arXiv:2211.09190}}.

\bibitem{Attems:2016tby}
M.~Attems, J.~Casalderrey-Solana, D.~Mateos, D.~Santos-Oliv\'an, C.~F.
  Sopuerta, M.~Triana, and M.~Zilh\~ao, {\it {Holographic Collisions in
  Non-conformal Theories}},  {\em JHEP} {\bf 01} (2017) 026,
  [\href{http://arxiv.org/abs/1604.06439}{{\tt arXiv:1604.06439}}].

\bibitem{Attems:2017zam}
M.~Attems, J.~Casalderrey-Solana, D.~Mateos, D.~Santos-Oliv\'an, C.~F.
  Sopuerta, M.~Triana, and M.~Zilh\~ao, {\it {Paths to equilibrium in
  non-conformal collisions}},  {\em JHEP} {\bf 06} (2017) 154,
  [\href{http://arxiv.org/abs/1703.09681}{{\tt arXiv:1703.09681}}].

\bibitem{Attems:2018gou}
M.~Attems, Y.~Bea, J.~Casalderrey-Solana, D.~Mateos, M.~Triana, and
  M.~Zilh\~ao, {\it {Holographic Collisions across a Phase Transition}},  {\em
  Phys. Rev. Lett.} {\bf 121} (2018), no.~26 261601,
  [\href{http://arxiv.org/abs/1807.05175}{{\tt arXiv:1807.05175}}].

\bibitem{Casalderrey-Solana:2016xfq}
J.~Casalderrey-Solana, D.~Mateos, W.~van~der Schee, and M.~Triana, {\it
  {Holographic heavy ion collisions with baryon charge}},  {\em JHEP} {\bf 09}
  (2016) 108, [\href{http://arxiv.org/abs/1607.05273}{{\tt arXiv:1607.05273}}].

\bibitem{Emparan:2013moa}
R.~Emparan, R.~Suzuki, and K.~Tanabe, {\it {The large D limit of General
  Relativity}},  {\em JHEP} {\bf 06} (2013) 009,
  [\href{http://arxiv.org/abs/1302.6382}{{\tt arXiv:1302.6382}}].

\bibitem{Emparan:2020inr}
R.~Emparan and C.~P. Herzog, {\it {Large D limit of Einstein\textquoteright{}s
  equations}},  {\em Rev. Mod. Phys.} {\bf 92} (2020), no.~4 045005,
  [\href{http://arxiv.org/abs/2003.11394}{{\tt arXiv:2003.11394}}].

\bibitem{Emparan:2014cia}
R.~Emparan and K.~Tanabe, {\it {Universal quasinormal modes of large D black
  holes}},  {\em Phys. Rev. D} {\bf 89} (2014), no.~6 064028,
  [\href{http://arxiv.org/abs/1401.1957}{{\tt arXiv:1401.1957}}].

\bibitem{Emparan:2014aba}
R.~Emparan, R.~Suzuki, and K.~Tanabe, {\it {Decoupling and non-decoupling
  dynamics of large D black holes}},  {\em JHEP} {\bf 07} (2014) 113,
  [\href{http://arxiv.org/abs/1406.1258}{{\tt arXiv:1406.1258}}].

\bibitem{Emparan:2015hwa}
R.~Emparan, T.~Shiromizu, R.~Suzuki, K.~Tanabe, and T.~Tanaka, {\it {Effective
  theory of Black Holes in the 1/D expansion}},  {\em JHEP} {\bf 06} (2015)
  159, [\href{http://arxiv.org/abs/1504.06489}{{\tt arXiv:1504.06489}}].

\bibitem{Emparan:2013xia}
R.~Emparan, D.~Grumiller, and K.~Tanabe, {\it {Large-D gravity and low-D
  strings}},  {\em Phys. Rev. Lett.} {\bf 110} (2013), no.~25 251102,
  [\href{http://arxiv.org/abs/1303.1995}{{\tt arXiv:1303.1995}}].

\bibitem{Emparan:2013oza}
R.~Emparan and K.~Tanabe, {\it {Holographic superconductivity in the large D
  expansion}},  {\em JHEP} {\bf 01} (2014) 145,
  [\href{http://arxiv.org/abs/1312.1108}{{\tt arXiv:1312.1108}}].

\bibitem{Emparan:2015rva}
R.~Emparan, R.~Suzuki, and K.~Tanabe, {\it {Quasinormal modes of (Anti-)de
  Sitter black holes in the 1/D expansion}},  {\em JHEP} {\bf 04} (2015) 085,
  [\href{http://arxiv.org/abs/1502.02820}{{\tt arXiv:1502.02820}}].

\bibitem{Emparan:2019obu}
R.~Emparan and R.~Suzuki, {\it {Topology-changing horizons at large D as Ricci
  flows}},  {\em JHEP} {\bf 07} (2019) 094,
  [\href{http://arxiv.org/abs/1905.01062}{{\tt arXiv:1905.01062}}].

\bibitem{Emparan:2014jca}
R.~Emparan, R.~Suzuki, and K.~Tanabe, {\it {Instability of rotating black
  holes: large D analysis}},  {\em JHEP} {\bf 06} (2014) 106,
  [\href{http://arxiv.org/abs/1402.6215}{{\tt arXiv:1402.6215}}].

\bibitem{Emparan:2015gva}
R.~Emparan, R.~Suzuki, and K.~Tanabe, {\it {Evolution and End Point of the
  Black String Instability: Large D Solution}},  {\em Phys. Rev. Lett.} {\bf
  115} (2015), no.~9 091102, [\href{http://arxiv.org/abs/1506.06772}{{\tt
  arXiv:1506.06772}}].

\bibitem{Emparan:2016sjk}
R.~Emparan, K.~Izumi, R.~Luna, R.~Suzuki, and K.~Tanabe, {\it {Hydro-elastic
  Complementarity in Black Branes at large D}},  {\em JHEP} {\bf 06} (2016)
  117, [\href{http://arxiv.org/abs/1602.05752}{{\tt arXiv:1602.05752}}].

\bibitem{Rozali:2017bll}
M.~Rozali, E.~Sabag, and A.~Yarom, {\it {Holographic Turbulence in a Large
  Number of Dimensions}},  {\em JHEP} {\bf 04} (2018) 065,
  [\href{http://arxiv.org/abs/1707.08973}{{\tt arXiv:1707.08973}}].

\bibitem{Andrade:2018nsz}
T.~Andrade, R.~Emparan, and D.~Licht, {\it {Rotating black holes and black bars
  at large D}},  {\em JHEP} {\bf 09} (2018) 107,
  [\href{http://arxiv.org/abs/1807.01131}{{\tt arXiv:1807.01131}}].

\bibitem{Andrade:2018rcx}
T.~Andrade, R.~Emparan, and D.~Licht, {\it {Charged rotating black holes in
  higher dimensions}},  {\em JHEP} {\bf 02} (2019) 076,
  [\href{http://arxiv.org/abs/1810.06993}{{\tt arXiv:1810.06993}}].

\bibitem{Andrade:2018yqu}
T.~Andrade, R.~Emparan, D.~Licht, and R.~Luna, {\it {Cosmic censorship
  violation in black hole collisions in higher dimensions}},  {\em JHEP} {\bf
  04} (2019) 121, [\href{http://arxiv.org/abs/1812.05017}{{\tt
  arXiv:1812.05017}}].

\bibitem{Andrade:2018zeb}
T.~Andrade, C.~Pantelidou, and B.~Withers, {\it {Large D holography with metric
  deformations}},  {\em JHEP} {\bf 09} (2018) 138,
  [\href{http://arxiv.org/abs/1806.00306}{{\tt arXiv:1806.00306}}].

\bibitem{Andrade:2019edf}
T.~Andrade, R.~Emparan, D.~Licht, and R.~Luna, {\it {Black hole collisions,
  instabilities, and cosmic censorship violation at large $D$}},  {\em JHEP}
  {\bf 09} (2019) 099, [\href{http://arxiv.org/abs/1908.03424}{{\tt
  arXiv:1908.03424}}].

\bibitem{Andrade:2019rpn}
T.~Andrade, C.~Pantelidou, J.~Sonner, and B.~Withers, {\it {Driven black holes:
  from Kolmogorov scaling to turbulent wakes}},  {\em JHEP} {\bf 07} (2021)
  063, [\href{http://arxiv.org/abs/1912.00032}{{\tt arXiv:1912.00032}}].

\bibitem{Andrade:2020ilm}
T.~Andrade, R.~Emparan, A.~Jansen, D.~Licht, R.~Luna, and R.~Suzuki, {\it
  {Entropy production and entropic attractors in black hole fusion and
  fission}},  {\em JHEP} {\bf 08} (2020) 098,
  [\href{http://arxiv.org/abs/2005.14498}{{\tt arXiv:2005.14498}}].

\bibitem{Emparan:2021ewh}
R.~Emparan, D.~Licht, R.~Suzuki, M.~Toma\v{s}evi\'c, and B.~Way, {\it {Black
  tsunamis and naked singularities in AdS}},  {\em JHEP} {\bf 02} (2022) 090,
  [\href{http://arxiv.org/abs/2112.07967}{{\tt arXiv:2112.07967}}].

\bibitem{Chihuahua-2022}
R.~Luna and M.~Sanchez-Garitaonandia, {\it
  \href{https://github.com/raimonluna/Chihuahua}{Chihuahua}},  2022.

\bibitem{KS_vs_physical}
V.~Latora and M.~Baranger, {\it Kolmogorov-sinai entropy rate versus physical
  entropy},  {\em Phys. Rev. Lett.} {\bf 82} (Jan, 1999) 520--523.

\bibitem{Kunihiro:2008gv}
T.~Kunihiro, B.~Muller, A.~Ohnishi, and A.~Schafer, {\it {Towards a Theory of
  Entropy Production in the Little and Big Bang}},  {\em Prog. Theor. Phys.}
  {\bf 121} (2009) 555--575, [\href{http://arxiv.org/abs/0809.4831}{{\tt
  arXiv:0809.4831}}].

\bibitem{Maldacena:2015waa}
J.~Maldacena, S.~H. Shenker, and D.~Stanford, {\it {A bound on chaos}},  {\em
  JHEP} {\bf 08} (2016) 106, [\href{http://arxiv.org/abs/1503.01409}{{\tt
  arXiv:1503.01409}}].

\bibitem{DMM}
D.~Ramirez, M.~Sanchez-Garitaonandia, and M.~Toma{\v s}evi\'c, {\it {The large
  $D$ limit of chaos}},  {\em To appear}.

\bibitem{Bea:2021ieq}
Y.~Bea, J.~Casalderrey-Solana, T.~Giannakopoulos, D.~Mateos,
  M.~Sanchez-Garitaonandia, and M.~Zilh\~ao, {\it {Domain collisions}},  {\em
  JHEP} {\bf 06} (2022) 025, [\href{http://arxiv.org/abs/2111.03355}{{\tt
  arXiv:2111.03355}}].

\end{thebibliography}\endgroup
\bibliographystyle{JHEP}
\end{document}